\documentclass[%
 reprint,
 amsmath,amssymb,
 aps,
]{revtex4-1}

\usepackage{graphicx}
\usepackage{dcolumn}
\usepackage{bm}
\usepackage{amsmath}
\usepackage{breqn}

\begin{document}


\title{New Solutions To The Bianchi IX Wheeler DeWitt Equation And Leading Order Solutions For $\Lambda$ $\ne$ 0 And A Primordial Magnetic Field}

\author{Daniel Berkowitz}
 \altaffiliation{Physics Department, Yale University.\\ daniel.berkowitz@yale.edu \\ This work is in memory of my parents, Susan Orchan Berkowitz, and Jonathan Mark Berkowitz}

\date{\today}

\begin{abstract}
New closed form solutions to the Lorentzian signature symmetry reduced Bianchi IX Wheeler DeWitt equation are obtained using a Euclidean-signature semi classical method for Hartle Hawking ordering parameters $ \pm 2\sqrt{33} $ for the Moncrief-Ryan (or ‘wormhole’) Hamilton-Jacobi solution. Moving past the wormhole case we also compute first order quantum corrections to Bianchi IX wave functions restricted to the $\beta_+$ axis for the 'no boundary' solution\cite{hartle1983wave} and one of the 'arm' solutions \cite{barbero1996minisuperspace}. Furthermore, six solutions to the Bianchi IX Euclidean-signature Hamiltonian Jacobi equation for the case when a cosmological constant is present have been found and are used to construct a semi-classical (leading order wave functions) which respects the symmetry of the Bianchi IX potential. Also two additional solutions were found when both a cosmological constant and an aligned primordial magnetic field\cite{uggla1995classifying} are present, which are also used to construct a leading order wave function. Furthermore we construct leading order 'excited' states which are restricted to the $\beta_+$ axis for the cases when a cosmological constant and a primordial magnetic field are present. To conclude, we discuss the interesting features of our solutions, and point out how the cosmological constant appears to act as a driver of anisotropy in our Bianchi IX models. The Euclidean-signature semi-classical method used here is applicable to certain \cite{moncrief2014euclidean} field theories as well. Because this semi classical method was able to find new solutions to a heavily studied problem the results in this paper represent significant progress in the Wheeler-DeWitt approach to quantum cosmology and in the application of Euclidean-signature semi classical methods to Lorentzian signature problems in general. 

\end{abstract}

\pacs{Valid PACS appear here}
\maketitle


\section{\label{sec:level1}INTRODUCTION}
Of all the Bianchi models, the Bianchi IX models have been the ones to be most thoroughly investigated. This is partly due to the interesting 'chaotic'\cite{barrow1982chaotic} \cite{chernoff1983chaos} \cite{cornish1997mixmaster} behavior they exhibit near their singularities and the fact that their spatial hyper-surfaces are a relatively intuitive topological 3-sphere. The topology of the Bianchi IX models is shared with the K=1 FLRW models, which previously was widely believed to be a good approximation for our physical universe, until more precise cosmological observations \cite{de2002multiple} showed otherwise. The vacuum diagonalized Bianchi IX models (or the Mixmaster models) were first studied by Belinsky, Khalatnikov and Lifshitz (BKL) to investigate the nature of cosmological singularities \cite{belinskii1970oscillatory}; and also by Misner, who used the Hamiltonian approach for the study of its dynamics and was the first to consider its quantization \cite{misner1969quantum} \cite{klauder1972magic}. Bianchi IX models have been studied in a plethora of contexts such as homogeneous relativistic cosmologies, supergravity \cite{ryan1975relativistic,graham1991supersymmetric,graham1993anisotropic,graham1993supersymmetric,bene1994supersymmetric,graham1994hartle,d1993quantization,csordas1995supersymmetric,damour2014quantum}, in Euclidean signature settings \cite{belinskii1978asymptotically}, and with local approximations to the full Bianchi IX potential \cite{marolf1995observables}.
The quantum cosmology of Bianchi IX models with a cosmological constant was previously investigated by employing Chern-Simons solutions in Ashtekar’s variables \cite{paternoga1996exact} \cite{graham1996physical} using a different class of operator ordering than the one we will exclusively consider. Classical Bianchi models with a primordial magnetic field have been investigated sporadically since the 1960s \cite{thorne1967primordial}. In the 80s Dieter Lorenz studied the classical Bianchi VIII and IX models \cite{lorenz1980exact} with matter and an electromagnetic field; other investigations into the Bianchi I \cite{leblanc1997asymptotic} and VI models \cite{leblanc1995asymptotic} were done in the 90s as well. In this paper we will only consider primordial magnetic fields from a diagonalized stress energy tensor whose contribution to the potential of the Hamiltonian has the form contained in \cite{uggla1995classifying}. To the author's knowledge there hasn't been any published investigations into quantum cosmological Bianchi A models with a primordial magnetic field. Besides what the author will present in this paper on the Bianchi IX models, he has already amassed a large amount of preliminary results for solutions to the Lorentzian signature symmetry reduced Wheeler Dewitt equations(for particular ordering parameters, and for arbitrary ordering parameters) for a variety of Bianchi A models with and without matter sources, in addition to their corresponding LRS models.

The first to apply the Euclidean-signature semi classical method to the Bianchi IX models and demonstrate that for the wormhole case smooth globally defined asymptotic solutions exist for arbitrary ordering parameter and formulate a way of computing those terms were Joseph Bae and Vincent Moncrief \cite{bae2015mixmaster} \cite{moncrief2014euclidean}.

The diagonalized Bianchi IX models we will quantize are parameterized by the following Misner variables $(\alpha,\beta_+,\beta_-)$ \cite{misner1969mixmaster}\cite{misner1969quantum}, where $\alpha(t)$ gives the size of the three-dimensional hyper surface relative to its initial size; the anisotropy parameters $(\beta(t)_+,\beta(t)_-)$ describe the anisotropy of the hyper surface or in other words how distorted the surface is from a perfectly round 3-sphere. All three Misner variables can range from $-\infty$ to $\infty$ and define the minisuperspace of the Bianchi IX models. Classically as mentioned the Bianchi IX models exhibit 'chaotic' behavior. This results in complications \cite{christiansen1995non} in trying to solve their symmetry reduced Wheeler DeWitt equation in Lorentian signature using traditional semi classical methods. 

To get around this and solve the Lorentzian signature Bianchi IX symmetry reduced Wheeler Dewitt equation we employ a Euclidean-signature semi classical method \cite{marini2019euclidean}\cite{moncrief2014euclidean} which converts the symmetry reduced Wheeler DeWitt equation into an infinite sequence of equations, the first among them being the Euclidean-signature Hamilton-Jacobi equation, followed by infinitely many linear transport equations. As will be shown in detail soon, this infinite sequence of equations can truncate if solutions exist for lower order transport equations which cause the solutions of the higher order transport equations to be satisfied by zero. If the sequence truncates then a closed form solution to the symmetry reduced Wheeler DeWitt equation can be written down. Even if the infinite sequence of equations cannot be terminated, under certain circumstances which are discussed in\cite{bae2015wormhole} and \cite{moncrief2014euclidean} one can prove inductively that smooth globally defined solutions exist for all of the higher order equations and thus establish an asymptotic solution. Furthermore, this method allows one to define 'ground' and 'excited' states \cite{moncrief2014euclidean}. These 'excited' states can be broken up into 'bound' and 'scattering' states, despite the physically relevant wave functions being those which are annihilated by the quantized Hamiltonian constraint $\hat{\mathcal{H}}_{\perp} \Psi=0$ and thus admitting only zero as eigenvalues. Because the symmetry reduced Wheeler DeWitt does not have the form of a Schr$\text{\" o}$dinger equation we must take with a grain of salt bestowing the classification of 'ground' or 'excited' onto these solutions. Nonetheless qualitatively the full solutions and leading order solutions which were found using this method by Joseph Bae in a previous work \cite{bae2015mixmaster} qualitatively behave like typical ground and excited states in ordinary quantum mechanics, so there is justification for calling them such which we will further elaborate on.

This paper is organized as follows: first, we review the canonical quantization of the diagonalized Bianchi IX models with a cosmological constant, outlining how we obtain the Hamiltonian constraint and the relevant symmetry reduced Wheeler-DeWitt equation. Next, we present the modified semi-classical method as developed by Moncrief et al. \cite{marini2019euclidean}, and the resulting equations it produces which in principle can be solved to construct an asymptotic solution to the symmetry reduced Wheeler DeWitt equation for arbitrary Hartle Hawking parameter. We will then show that only a finite number of them need to be solved to obtain closed form solutions for the Bianchi IX symmetry reduced Wheeler DeWitt equation for the factor ordering parameters $\pm 2\sqrt{33}$. Once the new closed form solutions generated directly from the semi-classical method are presented we will find more solutions using the rotational symmetry of the Bianchi IX potential. Before moving on we will discuss the matter of our closed form solutions not respecting the rotational symmetry of the potential, and use superposition to construct an interesting wave function out of the family of semi-classical(leading order) Bianchi IX 'wormhole' solutions that Joseph Bae\cite{bae2015mixmaster} computed. Afterwards we will compute first order quantum corrections to leading order wave functions restricted to the $\beta_+$ axis who's leading order (semi-classical) terms are the 'no boundary' solution, and one of the 'arm'\cite{barbero1996minisuperspace}\cite{bae2015wormhole} solutions. We will also construct leading order (semi-classical) 'excited' states for the wave functions for the aforementioned wave functions restricted to the $\beta_+$ axis.  In addition we will look at the "arm" solutions in the full $\beta$ plane, and plot a superposition of leading order wave functions composed of the "arm" solutions 'ground' and 'excited' 'wormhole' solutions, and the 'no boundary' solution. Then we will present solutions to the Euclidean-signature Hamilton-Jacobi equation for the Bianchi IX models with a cosmological constant, and an aligned primordial magnetic field, and use them to construct additional leading order solutions. The leading order solutions for the case with only a cosmological constant will respect the symmetry of the Bianchi IX potential after applying a straightforward group averaging procedure to them. We will then examine these solutions along the $\beta_+$ axis. Finally, we will discuss the interesting properties that these wave functions possess. It should be noted that past this point all references to the Wheeler DeWitt \cite{dewitt1967quantum} equation unless otherwise stated refer to the finite dimensional symmetry reduced Bianchi IX Wheeler DeWitt equation. Also we will use the terms leading order solution, and semi-classical solutions interchangeably.

\section{\label{sec:level1}The Bianchi IX Models With Cosmological Constant}

The Bianchi IX, or ‘Mixmaster’ cosmological models are spatially homogeneous, but anisotropic space-times defined on the manifold $R \times S^3$. Their metrics can be represented using the following basis whose elements are the one forms 

\begin{equation}
\begin{aligned}
&\sigma^1=\cos{\psi}d\theta+\sin{\psi}\sin{\theta}d\phi \\
&\sigma^2=\sin{\psi}d\theta-\cos{\psi}\sin{\theta}d\phi \\
&\sigma^3=d\psi+\cos{\theta}d\phi
\end{aligned}
\end{equation}

which satisfy 

\begin{equation}
d\sigma^i=\frac{1}{2}\epsilon_{ijk}\sigma^j\sigma^k
\end{equation}
where $ \epsilon_{ijk} $ is completely anti-symmetric with $\epsilon_{123} $  = 1.
When the stress energy tensor vanishes the non diagonal Bianchi IX line element can be expressed with a proper choice of gauge as a diagonal line element in terms of the one forms $\left(1\right)$, thus without loss of generality due to gauge equivalence we can write out the line element for the Bianchi IX models as the following

\begin{equation}
ds^2=-N^2dt^2+\frac{L^2}{6\pi}e^{2\alpha}(e{^{2\beta})_{ij}}\sigma^i\sigma^j
\end{equation}

where the space-time coordinates are $\left(t,\theta,\phi,\psi\right)$ and 
\begin{equation}
\left(e^{2 \beta}\right)=\operatorname{diag}\left(e^{2 \beta_{+}+2 \sqrt{3} \beta_{-}}, e^{2 \beta_{+}-2 \sqrt{3} \beta_{-}}, e^{-4 \beta_{+}}\right)
\end{equation}

which is expressed in Misner variables $\left(\alpha,\beta_+,\beta_-\right)$; $L$ can be any real number except zero and has dimensions of length which indicates in what unit $\alpha$ measures the spatial size of the hypersurface. To ensure that the metric is homogeneous the Misner variables can only depend on an evolution parameter. Furthermore to ensure that our metric has the proper signature, the lapse N$\left(t\right)$ cannot change sign or vanish for any real value of the evolution parameter.

In terms of Newton's constant, G, and the speed of light, c, the Einstein-Hilbert action for the case where the stress energy tensor vanishes is

\begin{equation}
I_{EH}=\int\frac{c^3}{16\pi G}\sqrt{-g}\\(R-2\Lambda)d^4x
\end{equation}
where $\sqrt{-g}$ is the square root of minus the determinant of the space-time metric, R is the Ricci scalar of the space-time metric and $d^4$x is dtd$ \theta $d$\phi$d$\psi$. The metric tensor $g_{uv}$ can be read off from the line element $\left(3\right)$. To obtain the Lagrangian we integrate over the spatial portion of the action, where our spatial variables will have the following ranges, $\theta$ $\in$ $\left(0,\pi\right)$, $\phi$ $\in$ $\left(0,2\pi\right)$, $\psi$ $\in$ $\left(0,4\pi\right)$. 

In addition because our variation with respect to the Misner variables and $\dot{\alpha}$ vanishes at the boundary points, we drop all terms which can be represented as total time derivatives. Thus, our Lagrangian \cite{arnowitt1962gravitation} \cite{misner1973freeman} can easily be read off from the resulting action.

\begin{equation}
\begin{aligned} I_{\mathrm{ADM}} & :=\frac{c^{3} L^{3} \pi}{G(6 \pi)^{3 / 2}} \int_{I} d t \Big\{\frac{6 e^{3 \alpha}}{N}\left(-\dot{\alpha}^{2}+\dot{\beta}_{+}^{2}+\dot{\beta}_{-}^{2}\right) \\ &-\frac{(6 \pi) N e^{\alpha}}{2 L^{2}}[e^{-8 \beta_{+}}-4 e^{-2 \beta_{+}} \cosh \left(2 \sqrt{3} \beta_{-}\right)\\ &+2  e^{4 \beta_{+}}\left(\cosh \left(4 \sqrt{3} \beta_{-}\right)-1\right)-\frac{2e^{2\alpha}L^2\Lambda}{3\pi}  \Big] \Big\}\\ & :=\int_{I} L_{\mathrm{ADM}} d t \end{aligned} 
\end{equation}

The Hamiltonian is constructed via applying the following Legendre transformation
\begin{equation}
\begin{aligned} 
\mathcal{H}_{\perp} & := p_{\alpha} \dot{\alpha}+p_{+} \dot{\beta}_{+}+p_{-} \dot{\beta}_{-} -L_{\mathrm{ADM}} \\
p_{\alpha}  & =\frac{\partial L_{\mathrm{ADM}}}{\partial \dot{\alpha}}=\frac{-c^{3} L^{3} \pi}{G(6 \pi)^{3 / 2}} \frac{12 e^{3 \alpha} \dot{\alpha}}{N} \\ p_{+}  & =\frac{\partial L_{\mathrm{ADM}}}{\partial \dot{\beta}_{+}}=\frac{c^{3} L^{3} \pi}{G(6 \pi)^{3 / 2}} \frac{12 e^{3 \alpha} \dot{\beta}_{+}}{N} \\ p_{-}  & =\frac{\partial L_{\mathrm{ADM}}}{\partial \dot{\beta}_{-}}=\frac{c^{3} L^{3} \pi}{G(6 \pi)^{3 / 2}} \frac{12 e^{3 \alpha} \dot{\beta}_{-}}{N} \end{aligned}
\end{equation}

In terms of these canonical variables $(p_{\alpha},p_{+},p_{-},\alpha,\beta_+,\beta_-)$ the Hamiltonian constraint for which the lapse acts as a Lagrange multiplier is the following. 

\begin{equation}
\begin{aligned} \mathcal{H}_{\perp} : &=\frac{(6 \pi)^{1 / 2} G}{4 c^{3} L^{3} e^{3 \alpha}} \Bigg\{ \left(-p_{\alpha}^{2}+p_{+}^{2}+p_{-}^{2}\right) \\ & + \left(\frac{c^{3}}{G}\right)^{2} L^{4} e^{4 \alpha} \Biggl[\frac{e^{-8 \beta_{+}}}{3}-\frac{4 e^{-2 \beta_{+}}}{3} \cosh \left(2 \sqrt{3} \beta_{-}\right) \\& +  \frac{2}{3} e^{4 \beta_{+}}\left(\cosh \left(4 \sqrt{3} \beta_{-}\right)-1 \right)+\frac{2e^{2\alpha}L^2\Lambda}{9\pi} \Biggr] \Bigg\}  
\end{aligned}
\end{equation}

To quantize the above classical system we will use standard canonical quantization. 

\begin{equation}
\begin{aligned} p_{\alpha} \longrightarrow \hat{p}_{\alpha} & :=\frac{\hbar}{i} \frac{\partial}{\partial \alpha} \\ p_{+} \longrightarrow \hat{p}_{+} & :=\frac{\hbar}{i} \frac{\partial}{\partial \beta_{+}} \\ p_{-} \longrightarrow \hat{p}_{-} & :=\frac{\hbar}{i} \frac{\partial}{\partial \beta_{-}} \end{aligned}
\end{equation}

Here $\hbar$=$\frac{h}{2\pi}$ where h is Planck's constant. 

Due to the ambiguity present in quantizing products of classical variables we will proceed to quantize the product of generalized coordinates and momenta using the above procedure by introducing the Hartle Hawking ordering parameter B \cite{hartle1983wave} in the following way. 

\begin{equation}
\begin{array}{l}{-e^{-3 \alpha} p_{\alpha}^{2} \longrightarrow \frac{\hbar^{2}}{e^{(3-B) \alpha}} \frac{\partial}{\partial \alpha}\left(e^{-B \alpha} \frac{\partial}{\partial \alpha}\right)} \\ {e^{-3 \alpha} p_{+}^{2} \longrightarrow \frac{-\hbar^{2}}{e^{3 \alpha}} \frac{\partial^{2}}{\partial \beta_{+}^{2}}} \\ {e^{-3 \alpha} p_{-}^{2} \longrightarrow \frac{-\hbar^{2}}{e^{3 \alpha}} \frac{\partial^{2}}{\partial \beta_{-}^{2}}}\end{array}
\end{equation}

In this work we assume that B is an arbitrary real number. 

If we substitute all of the above and do a little rearranging we obtain the Wheeler DeWitt equation which will be the main focus of this work
\begin{equation}
\begin{aligned}
& \left(\frac{\ell_{p}}{L}\right)^3  e^{-(3-B)\alpha}\frac{\partial}{\partial \alpha}\left(e^{-B\alpha}\frac{\partial \Psi}{\partial \alpha}\right)-e^{-3\alpha}\left(\frac{\partial^2 \Psi}{\partial \beta^2_{+}}+\frac{\partial^2 \Psi}{\partial \beta^2_{-}}\right) \\ & + \left(\frac{L}{\ell_{p}}\right)e^{\alpha}\Bigl(\frac{e^{-8\beta_+}}{3}-\frac{4}{3}e^{-2\beta_+}\cosh\left({2\sqrt{3}\beta_-}\right) \\ & + \frac{2}{3}e^{4\beta_+}\left(\cosh\left({4\sqrt{3}\beta_-}\right)-1\right)+\frac{2e^{2\alpha}L^2 \Lambda}{9\pi}\Bigr)\Psi=0, \\ &  \\ &
 \hat{\mathcal{H}}_{\perp} \Psi=0.
\end{aligned}
\end{equation}

When we include our aligned primordial magnetic field we will add on to the potential term of the above equation $2b^{2}e^{2\alpha-4\beta_+}$ \cite{uggla1995classifying}\cite{wainwright2005dynamical}

The above Wheeler DeWitt equation is the analogue to the time dependent Schr$\text{\" o}$dinger equation of quantum mechanics for Bianchi IX quantum cosmology. Viewing the Wheeler DeWitt equation as $\hat{\mathcal{H}}_{\perp} \Psi=0$, and trying to relate it to the conventional Schr$\text{\" o}$dinger equation results in the problem of time manifesting as
\begin{equation}
i \hbar \frac{\partial \Psi}{\partial t}=N \hat{\mathcal{H}}_{\perp} \Psi=0
\end{equation}
 where $\frac{\partial \Psi}{\partial t}=0$. Due to the absence of the time derivative term of the Schr$\text{\" o}$dinger equation in the Wheeler DeWitt equation, the construction of a unitary time evolution operator is not trivial, thus leading to the potential breakdown of a simple probabilistic interpretation of the wave function of the universe because $|\psi|^2$ is not conserved in any of the Misner variables. A Klein-Gordon current 
 \begin{equation}
\mathcal{J}=\frac{i}{2}\left(\psi^{*} \nabla \psi-\psi \nabla \psi^{*}\right)
\end{equation}
can be defined \cite{vilenkin1989interpretation} \cite{mostafazadeh2004quantum} which could be used to construct a probability density. It however possesses the unattractive features that it vanishes when the wave function used to construct the current is purely real or imaginary, and that it doesn't yield a positive definite quantity like $|\psi|^2$. With our closed form solutions which we will present shortly, complex linear combinations can be constructed which yield non trivial Klein-Gordon currents which can be used to extrapolate physics from these wave functions. Besides the issue of constructing a probability density function, it appears the quanitized Hamiltonian constraint admits only zero's as eigenvalues. This may lead one to the conclusion that all of the states which satisfy the Wheeler DeWitt equation possess vanishing energy. This on the surface makes it impossible to distinguish between ground and excited states because all states seemingly have the same energy. This apparent obstacle to delineate ground and excited states can be overcome by examining the nuanced nature of the ADM formalism \cite{arnowitt1959dynamical}. When cast in the ADM formalism general relativity is a constrained theory with four Lagrange multipliers, the lapse and the three components of the shift. The constraint associated with the lapse is due to general relativity being invariant under reparameterization of the evolution parameter, likewise the constraint associated with the shift is due to diffeomorphism invariance and is called the diffeomorphism constraint. The diffeomorphism constraint is due to the configuration space $h_ {ab} $ being too large. To remedy this one can define a superspace \cite{fischer1970theory} \cite{giulini2009superspace} where an equivalence class for $h_ {ab} $ is constructed such that two $h_ {ab} $ are in the same class if they can be carried into one another by a diffeomorphism. This shrinks the configuration space, allowing the diffeomorphism constraint to be satisfied. The same cannot be done for the reparameterization constraint \cite{wald1984general}. This explains why it wouldn't even make sense for a time derivative to be present because there is no unique "time" to use and partially explains the origins of the "problem of time". To get a better feel for what is going on one can examine the vanishing Hamiltonian of a fully constrained system. One can formulate the Lagrangian of a free particle moving in one dimension, and introduce another configuration variable by defining function t(T) where T is some arbitrary evolution parameter. If one were to treat both X(t(T)) and t(T) as configuration variables and formulate the system's Hamiltonian, they would notice that the Hamiltonian vanishes. Obviously the energy of a free one dimensional particle moving at a particular velocity cannot be zero. This is resolved by realizing that the dynamics of the system is now encoded in how X(t(T)) evolves with respect to t(T) where both are configuration variables. For an explicit demonstration of the above vanishing Hamiltonian construction we refer the reader to \cite{rovelli2014covariant}. In other words, for these types of constrained systems the Hamiltonian no longer corresponds to the total energy. Thus the Hamiltonian constraint we quantized does not represent the total energy of a space-time in general relativity, and its vanishing eigenvalues do not mean that only states which possess zero energy are physically allowed. This allows leeway in defining 'ground' and 'excited' states in which features of ordinary quantum mechanics manifest as will be demonstrated in the next section. A more in depth discussion in regards to how the Euclidean-signature semi classical method can be used to define 'ground' and 'excited' states despite them both being annihilated by the quantized Hamiltonian constraint can be found in \cite{moncrief2014euclidean}. To deal with the problem of time we will choose one of the Misner variables to be our "time" \cite{dewitt1967quantum}. A good time parameter increases monotonically, and out of the variables we can choose from $\alpha$ is the best candidate for "time". There is one catch. Classically a Bianchi IX universe starts out expanding but reaches a maximum size and starts to recollapse \cite{lin1989proof} \cite{lin1990proof} \cite{rendall1997global} \cite{ringstrom2001bianchi}. During the recollapsing epoch $\alpha$ decreases monotonically and "time" can be thought of as going in reverse. If we were to add an extra degree of freedom from a free scalar field, we can construct a truly monotonically increasing "clock" by denoting our "time" variable to be the scalar field. 
 
 \section{\label{sec:level1}The Euclidean-signature semi classical method} 
Our outline of this method will follow closely \cite{moncrief2014euclidean}. The method outlined in this section and its resultant equations can in principle be used to construct solutions (closed form and asymptotic) to a wide class of quantum cosmological models such as all of the Bianchi A models and their corresponding locally rotationally symmetric (LRS) models expressed in Misner variables. 
 
The first step we will take in solving the Wheeler DeWitt equation is to introduce the ansatz
 \begin{equation}
\stackrel{(0)}{\Psi}_{\hbar}=e^{-S_{\hbar} / \hbar}
\end{equation}

where $-S_{\hbar}$ is a function of $\left(\alpha,\beta_+,\beta_-\right)$ We will rescale $-S_{\hbar}$ in the following way  
\begin{equation}
\mathcal{S}_{\hbar} :=\frac{G}{c^{3} L^{2}} S_{\hbar}
\end{equation}

where $\mathcal{S}_{\hbar}$ is dimensionless and admits the following power series in terms of this dimensionless parameter
\begin{equation}
X :=\frac{L_{\text { Planck }}^{2}}{L^{2}}=\frac{G \hbar}{c^{3} L^{2}}.
\end{equation}

The series is given by 
\begin{equation}
\mathcal{S}_{\hbar}=\mathcal{S}_{(0)}+X \mathcal{S}_{(1)}+\frac{X^{2}}{2 !} \mathcal{S}_{(2)}+\cdots+\frac{X^{k}}{k !} \mathcal{S}_{(k)}+\cdots
\end{equation},

and as a result our initial ansatz now takes the following form 
\begin{equation}
\stackrel{(0)}{\Psi}_{\hbar}=e^{-\frac{1}{X} \mathcal{S}_{(0)}-\mathcal{S}_{(1)}-\frac{X}{2 !} \mathcal{S}_{(2)}-\cdots}
\end{equation}.

Substituting this ansatz into the Wheeler-DeWitt equation and
requiring satisfaction, order-by-order in powers of X leads immediately to the sequence of equations

\begin{equation}
\begin{aligned}
&{\left(\frac{\partial \mathcal{S}_{(0)}}{\partial \alpha}\right)^{2}-\left(\frac{\partial \mathcal{S}_{(0)}}{\partial \beta_{+}}\right)^{2}-\left(\frac{\partial \mathcal{S}_{(0)}}{\partial \beta_{-}}\right)^{2}}+V=0,\\&
V=e^{4\alpha}\Bigr[\frac{e^{-8\beta_+}}{3}-\frac{4}{3}e^{-2\beta_+}\cosh{\left(2\sqrt{3}\beta_-\right)} \\ & + \frac{2}{3}e^{4\beta_+}\left(\cosh{\left(4\sqrt{3}\beta_-\right)}-1\right)+\frac{2e^{2\alpha}L^2 \Lambda}{9\pi}\Bigr]
\end{aligned}
\end{equation}
\begin{equation}
\begin{aligned}
& 2\left[\frac{\partial \mathcal{S}_{(0)}}{\partial \alpha} \frac{\partial \mathcal{S}_{(1)}}{\partial \alpha}-\frac{\partial \mathcal{S}_{(0)}}{\partial \beta_{+}} \frac{\partial \mathcal{S}_{(1)}}{\partial \beta_{+}}-\frac{\partial \mathcal{S}_{(0)}}{\partial \beta_{-}} \frac{\partial \mathcal{S}_{(1)}}{\partial \beta_{-}}\right] \\ & +B \frac{\partial \mathcal{S}_{(0)}}{\partial \alpha}-\frac{\partial^{2} \mathcal{S}_{(0)}}{\partial \alpha^{2}}+\frac{\partial^{2} \mathcal{S}_{(0)}}{\partial \beta_{+}^{2}}+\frac{\partial^{2} \mathcal{S}_{(0)}}{\partial \beta_{-}^{2}}=0,
\end{aligned}
\end{equation},
\begin{equation}
\begin{aligned}
& 2\left[\frac{\partial \mathcal{S}_{(0)}}{\partial \alpha} \frac{\partial \mathcal{S}_{(k)}}{\partial \alpha}-\frac{\partial \mathcal{S}_{(0)}}{\partial \beta_{+}} \frac{\partial \mathcal{S}_{(k)}}{\partial \beta_{+}}-\frac{\partial \mathcal{S}_{(0)}}{\partial \beta_{-}} \frac{\partial \mathcal{S}_{(k)}}{\partial \beta_{-}}\right] \\ & {+k\left[B \frac{\partial \mathcal{S}_{(k-1)}}{\partial \alpha}-\frac{\partial^{2} \mathcal{S}_{(k-1)}}{\partial \alpha^{2}}+\frac{\partial^{2} \mathcal{S}_{(k-1)}}{\partial \beta_{+}^{2}}+\frac{\partial^{2} \mathcal{S}_{(k-1)}}{\partial \beta_{-}^{2}}\right]} \\ & + \sum_{\ell=1}^{k-1} \frac{k !}{\ell !(k-\ell) !}\Biggr(\frac{\partial \mathcal{S}_{(\ell)}}{\partial \alpha} \frac{\partial \mathcal{S}_{(k-\ell)}}{\partial \alpha}-\frac{\partial \mathcal{S}_{(\ell)}}{\partial \beta_{+}} \frac{\partial \mathcal{S}_{(k-\ell)}}{\partial \beta_{+}} \\& - \frac{\partial \mathcal{S}_{(\ell)}}{\partial \beta_{-}} \frac{\partial \mathcal{S}_{(k-\ell)}}{\partial \beta_{-}}\Biggl) =0
\end{aligned}
\end{equation}

We will refer to $\mathcal{S}_{(0)}$ in our Wheeler DeWitt wave functions as the leading order term, which can be used to construct a semi-classical approximate solution to the Lorentzian signature Wheeler DeWitt equation, and call $\mathcal{S}_{(1)}$ the first order term. The $\mathcal{S}_{(1)}$ term can also be viewed as our first quantum correction, with the other $\mathcal{S}_{(k)}$ terms being additional higher order quantum corrections. Our closed form wave functions will not require terms beyond first order. 

If one can find a solution to the $\mathcal{S}_{(1)}$ equation which allows the $\mathcal{S}_{(2)}$ equation to be satisfied by zero, then one can write down the following as a solution to the Wheeler DeWitt equation for either a particular value of the Hartle-Hawking ordering parameter, or for an arbitrary ordering parameter depending on the properties of the $\mathcal{S}_{(1)}$ which is found.   

\begin{equation}
\stackrel{(0)}{\Psi}_{\hbar}=e^{-\frac{1}{X} \mathcal{S}_{(0)}-\mathcal{S}_{(1)}}
\end{equation}.

This can be easily shown. Lets take $\mathcal{S}_{(0)}$ and $\mathcal{S}_{(1)}$ as arbitrary known functions which allow the $\mathcal{S}_{(2)}$ transport equation to be satisfied by zero then the $k=3$ transport equation can be expressed as 
\begin{equation}
{2\left[\frac{\partial \mathcal{S}_{(0)}}{\partial \alpha} \frac{\partial \mathcal{S}_{(3)}}{\partial \alpha}-\frac{\partial \mathcal{S}_{(0)}}{\partial \beta_{+}} \frac{\partial \mathcal{S}_{(3)}}{\partial \beta_{+}}-\frac{\partial \mathcal{S}_{(0)}}{\partial \beta_{-}} \frac{\partial \mathcal{S}_{(3)}}{\partial \beta_{-}}\right]}=0
\end{equation}
which is clearly satisfied by $\mathcal{S}_{(3)}$=0. The $\mathcal{S}_{(4)}$ equation can be written in the same form and one of its solution is 0 as well, thus resulting in the $\mathcal{S}_{(5)}$ equation possessing the same form as (23). One can easily convince themselves that this pattern continues for all of the $k\geq 3$ $\mathcal{S}_{(k)}$ transport equations as long as the solution of the $\mathcal{S}_{(k-1)}$ transport equation is chosen to be 0. Thus in some situations a $\mathcal{S}_{(1)}$ exists which allows one to set the solutions to all of the higher order transport equations to zero, which results in the infinite sequence of transport equations generated by our ansatz to truncate to a finite sequence of equations whose solutions allow us to construct a closed form wave function satisfying the Wheeler DeWitt equation. Not all solutions to the $\mathcal{S}_{(1)}$ transport equation will allow the $\mathcal{S}_{(2)}$ transport equation to be satisfied by zero; however in our case, we were able to find $\mathcal{S}_{(1)}$'s which cause the $\mathcal{S}_{(2)}$ transport equation to be satisfied by zero, thus allowing one to set all of the solutions to the higher order transport equations to zero as shown above. This will enable us to construct closed form solutions to the Lorentzian signature Bianchi IX Wheeler Dewitt equation for particular Hartle Hawking ordering parameters.   

To calculate 'excited' states we introduce the following ansatz. 
\begin{equation}
{\Psi}_{\hbar}={\phi}_{\hbar} e^{-S_{\hbar} / \hbar}
\end{equation}
where $$
S_{\hbar}=\frac{c^{3} L^{2}}{G} \mathcal{S}_{\hbar}=\frac{c^{3} L^{2}}{G}\left(\mathcal{S}_{(0)}+X \mathcal{S}_{(1)}+\frac{X^{2}}{2 !} \mathcal{S}_{(2)}+\cdots\right)
$$
is the same series expansion as before and ${\phi}_{\hbar}$ can be expressed as the following series 
\begin{equation}
{\phi_{\hbar}=\phi_{(0)}+X \phi_{(1)}+\frac{X^{2}}{2 !} \phi_{(2)}+\cdots+\frac{X^{k(*)}}{k !} \phi_{(k)}+\cdots}
\end{equation}
with X being the same dimensionless quantity as before. 
Inserting $\left(24\right)$ with the expansions given by $\left(17\right)$ and $\left(25\right)$ into the Wheeler DeWitt equation and by matching equations in powers of X leads to the following series of equations. 
\begin{equation}
-\frac{\partial \phi_{(0)}}{\partial \alpha} \frac{\partial \mathcal{S}_{(0)}}{\partial \alpha}+\frac{\partial \phi_{(0)}}{\partial \beta_{+}} \frac{\partial \mathcal{S}_{(0)}}{\partial \beta_{+}}+\frac{\partial \phi_{(0)}}{\partial \beta_{-}} \frac{\partial \mathcal{S}_{(0)}}{\partial \beta_{-}}=0,
\end{equation},
\begin{equation}
\begin{aligned}
&{-\frac{\partial \phi_{(1)}}{\partial \alpha} \frac{\partial \mathcal{S}_{(0)}}{\partial \alpha}+\frac{\partial \phi_{(1)}}{\partial \beta_{+}} \frac{\partial \mathcal{S}_{(0)}}{\partial \beta_{+}}+\frac{\partial \phi_{(1)}}{\partial \beta_{-}} \frac{\partial \mathcal{S}_{(0)}}{\partial \beta_{-}}} \\ & {+\left(-\frac{\partial \phi_{(0)}}{\partial \alpha} \frac{\partial \mathcal{S}_{(1)}}{\partial \alpha}+\frac{\partial \phi_{(0)}}{\partial \beta_{+}} \frac{\partial \mathcal{S}_{(1)}}{\partial \beta_{+}}+\frac{\partial \phi_{(0)}}{\partial \beta_{-}} \frac{\partial \mathcal{S}_{(1)}}{\partial \beta_{-}}\right)} \\ & {+\frac{1}{2}\left(-B \frac{\partial \phi_{(0)}}{\partial \alpha}+\frac{\partial^{2} \phi_{(0)}}{\partial \alpha^{2}}-\frac{\partial^{2} \phi_{(0)}}{\partial \beta_{+}^{2}}-\frac{\partial^{2} \phi_{(0)}}{\partial \beta_{-}^{2}}\right)=0,}
\end{aligned}
\end{equation}
\begin{equation}
\begin{aligned}
& -\frac{\partial \phi_{(k)}}{\partial \alpha} \frac{\partial \mathcal{S}_{(0)}}{\partial \alpha}+\frac{\partial \phi_{(k)}}{\partial \beta_{+}} \frac{\partial \mathcal{S}_{(0)}}{\partial \beta_{+}}+\frac{\partial \phi_{(k)}}{\partial \beta_{-}} \frac{\partial \mathcal{S}_{(0)}}{\partial \beta_{-}} \\ & 
+k\Biggr(-\frac{\partial \phi_{(k-1)}}{\partial \alpha} \frac{\partial \mathcal{S}_{(1)}}{\partial \alpha}+\frac{\partial \mathcal{S}_{(1)}}{\partial \beta_{+}} \frac{\partial \mathcal{S}_{(1)}}{\partial \beta_{+}}+\frac{\partial \phi_{(k-1)}^{(*)}}{\partial \beta_{-}} \frac{\partial \mathcal{S}_{(1)}}{\partial \beta_{-}}\Biggr) \\ & 
+\frac{k}{2}\Biggr(-B \frac{\partial \phi_{(k-1)}}{\partial \alpha}+\frac{\partial^{2} \phi_{(k-1)}^{(*)}}{\partial \alpha^{2}}-\frac{\partial^{2} \phi_{(k-1)}}{\partial \beta_{+}^{2}}-\frac{\partial^{2} \phi_{(k-1)}}{\partial \beta_{-}^{2}}\Biggr)\\ & -
\sum_{\ell=2}^{k} \frac{k !}{\ell !(k-\ell) !}\Biggr( \frac{\partial \phi_{(k-\ell)}}{\partial \alpha} \frac{\partial \mathcal{S}_{(\ell)}}{\partial \alpha}-\frac{\partial \phi_{(k-\ell)}}{\partial \beta_{+}} \frac{\partial \mathcal{S}_{(\ell)}}{\partial \beta_{+}} - \\ &   \frac{\partial \phi_{(k-\ell)}}{\partial \beta_{-}} \frac{\partial \mathcal{S}_{(\ell)}}{\partial \beta_{-}}\Biggr) =0.
\end{aligned}
\end{equation}

As can be seen from computing $\frac{d\phi_{(0)}\left(\alpha,\beta_+,\beta_-\right)}{dt}=\dot{\alpha}\frac{\partial \phi_{(0)}}{\partial \alpha}+\dot{\beta_+}\frac{\partial \phi_{(0)}}{\partial \beta_+}+\dot{\beta_-}\frac{\partial \phi_{(0)}}{\partial \beta_-}$, and inserting $\left(4.9, \hspace{1 mm} 4.18-4.20\right)$ from \cite{moncrief2014euclidean} that $\phi_{(0)}$ is a conserved quantity under the flow of $S_{0}$. This means any function $F\left(\phi_{(0)}\right)$ is also a solution of equation $\left(26\right)$. Wave functions constructed from these functions of $\phi_{0}$ are only physical if they are smooth and globally defined. Beyond the semi-classical limit, if smooth globally defined solutions can be proven to exist for the higher order $\phi$ transport equations then one can construct a family of 'excited' states and take a superposition of them like in ordinary quantum mechanics. A plot of such a superposition for the 'ground' and 'excited' states of the Bianchi IX models is displayed in figure(3) and (4).

Our 'excited' states ${\Psi}_{\hbar}={\phi}_{\hbar} e^{-S_{\hbar} / \hbar}$ qualitatively possess the same form as excited states for the quantum harmonic oscillator $ \psi_{n}(x)=H_{n}\left(\sqrt{\frac{m \omega}{\hbar}} x\right)\frac{1}{\sqrt{2^{n} n !}}\left(\frac{m \omega}{\pi \hbar}\right)^{1 / 4}  e^{-\frac{m \omega x^{2}}{2 \hbar}}$, where $H_n$ are the Hermite polynomials in which n is a positive integer which specifies its form. Because the solutions of the $\phi_{(0)}$ equation are quantities conserved along the flow generated by $\mathcal{S}_{(0)}$, any multiple $\phi^{n}_{(0)}$ also satisfies equation $\left(26\right)$. On purely physical grounds the amount of numbers required to specify an 'excited' state equals the number of excitable degrees of freedom present. For Bianchi A models with non dynamical matter sources that amounts to two numbers corresponding to the two anistropic degrees of freedoms. As a result our $\phi_{(0)}$ which distinguishes our 'excited' states from 'ground' states has the following form $ \prod_{i = 1}^{n} f^{m_{i}}\left(\alpha,\beta_+,\beta_-\right)_{i}$; where $f\left(\alpha,\beta_+,\beta_-\right)_{i}$ are independent conserved quantities satisfying equation $\left(26\right)$ which are raised to the power $m_{i}$, and n is the number of excitable degrees of freedom. If all of the $f\left(\alpha,\beta_+,\beta_-\right)_{i}$'s vanish at some point or points in minisuperspace then to ensure that our wave function is smooth and globally defined we must restrict $m_{i}$ to be positive integers which results in our 'excited' states being 'bound' states just like the quantum harmonic oscillator. This discretization of the quantities that are used to denote our 'excited' states is the mathematical manifestation of quantization one would expect excited states to possess. If none of our conserved quantities $f\left(\alpha,\beta_+,\beta_-\right)_{i}$ vanish in minisuperspace than our 'excited' states are 'scattering' states akin to the quantum free particle and $m_{i}$ can be any real or complex number. It is also possible that only some $f\left(\alpha,\beta_+,\beta_-\right)_{i}$'s vanish while the other do not, in this case our 'excited' states are a hybrid of 'bound' and 'scattering' states which means some degrees of freedoms are 'bound' while other are 'scattering'. As the author will go into great detail to explain in another work these hybrid 'excited' states within the context of the Euclidean-signature semi classical method exist for the diagnolized quantum Bianchi VIII models. Additional information for why we can call the above 'excited' states despite them being solutions to an equation which does not have the same form as the Schr$\text{\" o}$dinger equation can be found in \cite{moncrief2014euclidean}. In what follows we will set $L=1$, $c=1$, $G=1$ and $\hbar=1$.  

It should be mentioned that there is a relation between the sequence of 'ground' state transport equations $\left(20-21\right)$ and $\left(26\right)$. If the source terms of equations $\left(20-21\right)$ vanish, then those equations become identical to $\left(26\right)$, and possess identical solutions which are conserved quantities along the flow generated by $\mathcal{S}_{(0)}$. In such a case one can proceed in four different ways. First one can set $\phi_{0}$ to unity and $\mathcal{S}_{( k \geq 1)}$ to zero and construct a closed form solution to the Wheeler DeWitt equation. Second, one can choose either $\phi_{0}$ or  $\mathcal{S}_{( k \geq 1)}$ to be trivial and the other a non trivial conserved quantity. From here one can solve the higher order transport equations associated with the term they chose to be their non trivial conserved quantity. Third, one can choose both $\phi_{0}$ and $\mathcal{S}_{( k \geq 1)}$ to be the same non trivial conserved quantity. From here one can continue to solve both sets of transport equations. Fourth, one can proceed similarly to our third choice but have  $\phi_{0}$ and $\mathcal{S}_{( k \geq 1)}$ not equal each other. In these cases the distinction between terms which contribute to the 'ground' state and terms which contribute to the 'excited' states becomes blurred. Using a natural looking ansatz which will be introduce explicitly later, we can ad-hoc restrain the form of the non-trivial conserved quantities we obtain from solving the homogeneous form of the transport equations $\left(20-21\right)$ and $\left(26\right)$.

Furthermore even if the source terms do not vanish one is free to add to the inhomogeneous solutions of equation $\left(20-21\right)$ their homogeneous solutions to form other solutions to the $\mathcal{S}_{(k)}$ equations. This means one can in principle begin constructing 'excited' states directly from 'ground' state transport equations $\left(20-21\right)$. This goes to show that we have to be careful how we 'denote' ground and 'excited' states within the context of this method. From our transport equations, 'excited' states could be seen as first order quantum corrections to 'ground' states.  To differentiate between 'ground' and 'excited' states we will ultimately appeal to their qualitative features such as having multiple local max/mins or peaks, and their mathematical manifestation of quantization or discreteness. As will be elaborated on in the next section, when the potential of the quantum cosmological model we are considering has a certain property, one can use a natural looking ansatz to solve the required 'ground' and 'excited' state transport equations.
\section{\label{sec:level1}Closed Form Wormhole Solutions For The Wheeler DeWitt Equation}
The known solutions to the Euclidean-signature vacuum Hamilton Jacobi equation with $\Lambda$=0 are the Moncrief-Ryan 'wormhole' solution \cite{moncrief1991amplitude} \cite{graham1991supersymmetric}, the Hartle Hawking no 'boundary solution' \cite{graham1993supersymmetric} \cite{bene1994supersymmetric}, and three "arm" solutions \cite{barbero1996minisuperspace} which are related to each other via $\frac{2\pi}{3}$ rotations in the $\beta$ plane and the aforementioned solutions multiplied by ($-1$). These solutions to the Hamilton Jacobi equation get their names from the geometry that they induce on the Euclidean signature Bianchi IX 'spacetimes' due to the flows in minisuperspace that they generate. For the equations below the $+$ sign indicates, respectively the Moncrief-Ryan wormhole, the Hartle Hawking no boundary and the arm solutions.
\begin{equation}
\mathcal{S}_{(0  \hspace{1 mm} wh)}:= \pm   \frac{1}{6} e^{2 \alpha}\left(e^{-4 \beta_{+}}+2 e^{2 \beta_{+}} \cosh \left(2 \sqrt{3} \beta_{-}\right)\right)
\end{equation}
\begin{equation}
\begin{aligned}
\mathcal{S}_{(0  \hspace{1 mm} nb)} := & \pm \frac{1}{6} e^{2 \alpha}\Bigl[\left(e^{-4 \beta_{+}}+2 e^{2 \beta_{+}} \cosh \left(2 \sqrt{3} \beta_{-}\right)\right)  \\ & -2\left(e^{2 \beta_{+}}+2 e^{-\beta_{+}} \cosh \left(\sqrt{3} \beta_{-}\right)\right)\Bigr]
\end{aligned}
\end{equation}
\begin{equation}
\begin{aligned}
\mathcal{S}_{(0 \hspace{1 mm} arm1)}:= & \pm \Bigl[ \Bigl( \frac{1}{6} \mathrm{e}^{2 \alpha+2 \beta_{+}}\left(\mathrm{e}^{-6 \beta_{+}}+2 \cosh\left( 2 \sqrt{3} \beta_{-}\right)\right) \\ & + \frac{1}{3} \mathrm{e}^{2 \alpha-\beta_{+}}\left(\mathrm{e}^{3 \beta_{+}}-2 \sinh \left( \sqrt{3} \beta_{-} \right )\right) \Bigr]
\end{aligned}
\end{equation}
\begin{equation}
\begin{aligned}
\mathcal{S}_{(0 \hspace{1 mm} arm2)}:= & \pm \Bigl[ \Bigl( \frac{1}{6} \mathrm{e}^{2 \alpha+2 \beta_{+}}\left(\mathrm{e}^{-6 \beta_{+}}+2 \cosh \left(2 \sqrt{3} \beta_{-}\right)\right) \\ & +\frac{1}{3} \mathrm{e}^{2 \alpha-\beta_{+}}\left(\mathrm{e}^{3 \beta_{+}}+2 \sinh \left( \sqrt{3} \beta_{-}\right)\right)  \Bigr]
\end{aligned}
\end{equation} 
\begin{equation}
\begin{aligned}
\mathcal{S}_{(0 \hspace{1 mm} arm3)}:= & \pm \Bigl[ \Bigl(\frac{1}{6} \mathrm{e}^{2 \alpha+2 \beta_{+}}\left(\mathrm{e}^{-6 \beta_{+}}+2 \cosh \left( 2 \sqrt{3} \beta_{-} \right) \right) \\ & +\frac{1}{3} \mathrm{e}^{2 \alpha-\beta_{+}}\left(-\mathrm{e}^{3 \beta_{+}}+2 \cosh \left( \sqrt{3} \beta_{-} \right)\right) \Bigr]
\end{aligned}
\end{equation}

Wave functions of the form $e^{-\frac{1}{X} \mathcal{S}_{(0)}-\mathcal{S}_{(1)}-\frac{X}{2 !} \mathcal{S}_{(2)}-\cdots}$ which contain the minus signed $\mathcal{S}_{(0)}$ terms are pathological because they grow without bound as functions of the $\beta$ variables. This results in wave functions which are not normalizable at a fixed $\alpha$. This mystifies their physical interpretation and introduces doubt whether they represent physical universes at all. Perhaps if they only appeared in the Bianchi IX case we could just sweep them under the rug. However Wheeler DeWitt wave functions which have this pathological behavior are present in all Bianchi A models. For example in the Bianchi VIII model the known solutions\cite{obregon1996psi} corresponding to Hartle Hawking ordering parameter $B=\pm 6$ suffer from this pathology no matter what sign their leading order term takes. As a result the author was compelled to include these terms for the sake of completeness, and encourages others or himself in the future to further look into the physicality of these wave functions constructed from the minus sign $\mathcal{S}_{(0)}$ terms. 
For the remainder of this section we shall exclusively only work with  $\mathcal{S}_{(0  \hspace{1 mm} wh)}$. In a previous investigation of the Bianchi IX models using the Euclidean-signature semi classical method \cite{bae2015mixmaster} \cite{moncrief2014euclidean} only globally defined solutions of the transport equations were used to construct solutions for the Wheeler DeWitt equation. For example the following solution 
\begin{equation}
\mathcal{S}_{(1)}=-\frac{1}{2}(B+6) \alpha
\end{equation}

to the $\mathcal{S}_{(1)}$ transport equation which Joseph Bae found makes the source terms of the $\mathcal{S}_{(2)}$ transport equation vanish for ordering parameters $B=\pm 6$, thus resulting in the subsequent transport equations being able to take the form of (23). However, if we relax the condition that the solutions $\mathcal{S}_{(k)}$ to the transport equations must be globally defined we can compute new globally defined solutions to the Wheeler DeWitt equation. We are allowed to do this because ultimately all we seek are globally defined wave functions and the exponential of certain non globally defined functions, such as $\log (\sinh (\beta_-))$, are globally defined. The form of the wave function we choose, namely  $\stackrel{(0)}{\Psi}_{\hbar}=e^{-\frac{1}{X} \mathcal{S}_{(0)}-\mathcal{S}_{(1)}-\frac{X}{2 !} \mathcal{S}_{(2)}-\cdots}$,

allows us to take advantage of the above fact. The $\mathcal{S}_{(1)}$ equation we wish to solve is the following. 

\begin{equation}
\begin{aligned}
& \Biggl(2 e^{6 \beta_+} \cosh \left(2 \sqrt{3} \beta_-\right)
   \left(2 \frac{\partial \mathcal{S}_{(1)}}{\partial \alpha}-2 \frac{\partial \mathcal{S}_{(1)}}{\partial \beta_+}+\text{B}+6\right) \\ & +2 \frac{\partial \mathcal{S}_{(1)}}{\partial \alpha}-4 \sqrt{3} e^{6 \beta_+} \frac{\partial \mathcal{S}_{(1)}}{\partial \beta_-} \sinh
   \left(2 \sqrt{3} \beta_-\right) +4 \frac{\partial \mathcal{S}_{(1)}}{\partial \beta_+} \\&+\text{B}+6\Biggr) =0
\end{aligned}
\end{equation}

To facilitate finding all possible $\mathcal{S}_{(1)}$s which allow the $\mathcal{S}_{(2)}$ equation to be satisfied by zero, we will try to construct an ansatz which includes as many non trivial free parameters as possible. The hope is that these parameters can be adjusted so that the source term of the $\mathcal{S}_{(2)}$ equation vanishes. The author advocates this approach for finding solutions to the $\mathcal{S}_{(1)}$ equation for any cosmological model. 

Because the $\alpha$ dependence factors out we are free seek solutions of the following form

\begin{equation}
\begin{aligned}
\mathcal{S}_{(1)}= x1\alpha +f\left(\beta_+,\beta_-\right)
\end{aligned}
\end{equation}

where x1 is a arbitrary number. Plugging this into the $\mathcal{S}_{(1)}$ transport equation yields. 

\begin{equation}
\begin{aligned}
& 2 e^{6 \beta_+} \cosh \left(2 \sqrt{3} \beta_-\right) \left(-2 \frac{\partial f}{\partial \beta_+}+\text{B}+2 \text{x1}+6\right) \\ &-4 \sqrt{3} e^{6 \beta_+} \frac{\partial f}{\partial \beta_-} \sinh \left(2 \sqrt{3} \beta_-\right) +4  \frac{\partial f}{\partial \beta_+}+\text{B}+2 \text{x1}+6=0
\end{aligned}
\end{equation}

This is a non homogeneous transport equation with variable coefficients which can be solved using a computer algebraic system such as Mathematica or by applying a change of variables to turn this equation into a homogeneous transport equation and applying standard textbook techniques. A solution to this equation is 
\begin{equation}
\begin{aligned}
f\left(\beta_+,\beta_-\right)=& \frac{1}{8} (\text{B}+2 \text{x1}+6) \biggl(\log \left(\frac{1}{2} \sinh \left(2 \sqrt{3} \beta_-\right)\right)\\ &-2 \beta_+\biggr)
\end{aligned}
 \end{equation}

where x1 is an arbitrary number, and B is an arbitrary real number. We will rewrite this solution by making the following substitution $X1=-\frac{1}{8}\left(B+2x1+6\right)$ resulting in the following for our $\mathcal{S}_{(1)}$ 
 
\begin{equation}
\mathcal{S}_{(1)}:=\frac{1}{2} \alpha (-\text{B}-8 \text{X1}-6)+\text{X1} \left(2 \beta_+-\log \left(\sinh \left(2 \sqrt{3} \beta_-\right)\right)\right)
\end{equation}

A key feature of this $\mathcal{S}_{(1)}$ quantum correction is that it has two free parameters x1 and B. This will allow us to fix them to particular values which will cause the source term in the $\mathcal{S}_{(2)}$ equation to vanish and subsequently allow all of the higher order transport equations to be satisfied by zero, thus leaving us with a closed form solution of the Wheeler DeWitt equation.

The values of X1 and B where the source term for the $\mathcal{S}_{(2)}$ transport equation vanishes are the following $\left(X1=0,B=-6\right)$, $\left(X1=0,B=6\right)$, $\left(X1=1,B=-2\sqrt{33}\right)$, and $\left(X1=1,B=2\sqrt{33}\right)$. The pair $(X1=0,B=-6)$ and $(X1=0,B=6)$ correspond to solutions which have been known since the 90s \cite{moncrief1991amplitude} and were re-derived using the Euclidean-signature semi classical method \cite{bae2015mixmaster} \cite{moncrief2014euclidean}; the new solutions correspond to the pair $\left(X1=1,B=-2\sqrt{33}\right)$, and $\left(X1=1,B=2\sqrt{33}\right)$. For the remainder of this section we will focus on solutions corresponding to  $\left(X1=1,B=2\sqrt{33}\right)$. The analysis which we will perform can be duplicated exactly for solutions corresponding to $\left(X1=1,B=-2\sqrt{33}\right)$. Our base solutions which we will use to derive other solutions are the following. 
\begin{equation}
\begin{aligned}
\psi_{\mp 1}= \sinh \left(2 \sqrt{3} \beta_-\right) e^{\left(7+\sqrt{33}\right) \alpha-2 \beta_+  \mp \mathcal{S}_{(0 \hspace{1 mm} wh)}} 
\end{aligned}
\end{equation}

Note that if X1 is not a positive integer a $\sinh ^{-\text{X1}}\left(2 \sqrt{3} b\right)$ term would be present in the wave function, which would cause it to be not smooth nor necessarily globally defined. 
Four additional solutions can be derived by rotating  $\psi_{\mp 1}(\alpha,\beta_+,\beta_-)$ in the $(\beta_+,\beta_-)$ plane by $\frac{2\pi}{3}$ and $\frac{4\pi}{3}$. 
\begin{equation}
\psi_{\mp 2}=\psi_{\mp 1}\left(\alpha,-\frac{\sqrt{3}\beta_-}{2}-\frac{\beta_+}{2},-\frac{\beta_-}{2}+\frac{\sqrt{3}\beta_+}{2}\right)
\end{equation}
\begin{equation}
\psi_{\mp 3}=\psi_{\mp 1}\left(\alpha,\frac{\sqrt{3}\beta_-}{2}-\frac{\beta_+}{2},-\frac{\beta_-}{2}-\frac{\sqrt{3}\beta_+}{2}\right)
\end{equation}
It should be noted that $\psi_{\mp 1}$,$\psi_{\mp 2} $, and $\psi_{\mp 3} $ are not independent of each other because $\psi_{\mp 2}$+$\psi_{\mp 3}$ $=-\psi_{\mp 1} $. Additional solutions can be constructed by taking arbitrary linear combinations of the above four independent solutions
\begin{equation}
\psi = a_1\psi_{-1}+a_2\psi_{-2}+a_3\psi_{+1}+a_4\psi_{+2}
\end{equation}
where $a_{i}$ can be any number, real, imaginary or complex. Two plots of our closed form solutions are shown below. 

\begin{figure}[!ht]
\begin{minipage}[c]{0.4\linewidth}
\includegraphics[scale=.1]{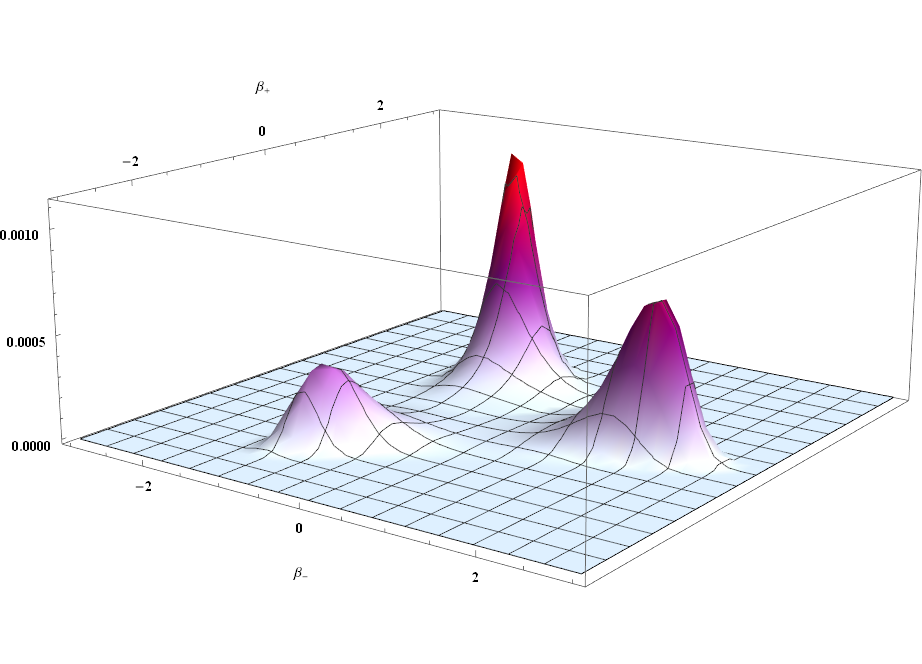}
\caption{$\alpha=-1$ $\hspace{1 mm} | \psi_{-1}-2\psi_{-2}|$}
\end{minipage}
\hfill
\begin{minipage}[c]{0.4\linewidth}
\includegraphics[scale=.1]{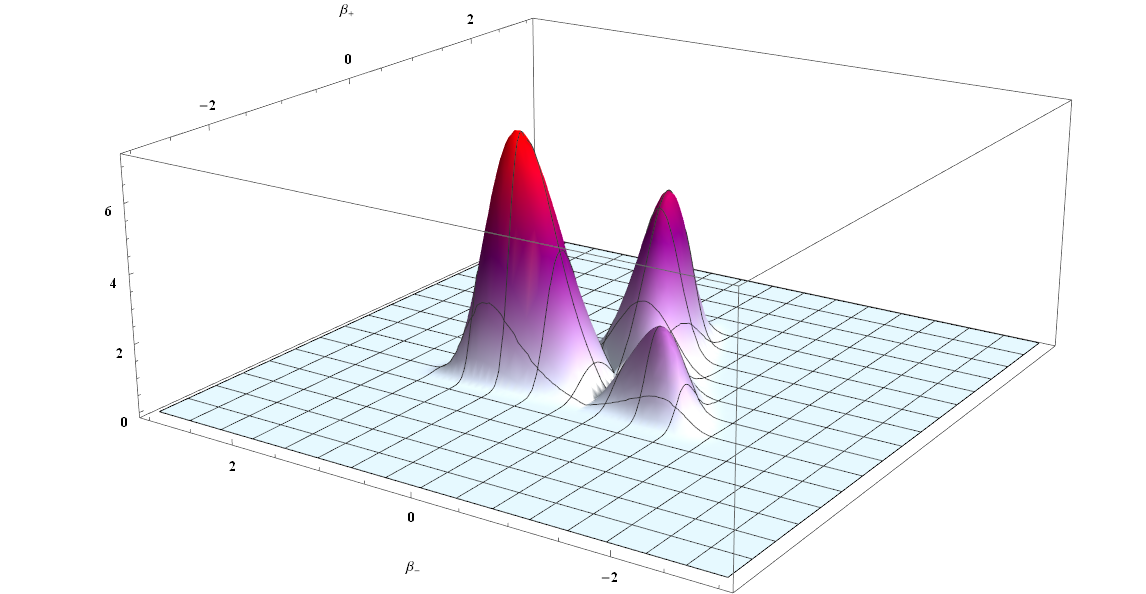}
\caption{$\alpha=1$ $\hspace{1 mm}| \psi_{-1}-2\psi_{-2}|$}
\end{minipage}%
\end{figure}

As one can see our wave functions which were found by exclusively using the 'ground' state transport equations posses the same form as an 'excited' state ${\Psi}_{\hbar}={\phi}_{\hbar} e^{-S_{\hbar} / \hbar}$. This is because when introducing our ansatz $\left( x_1\alpha +f\left(\beta_+,\beta_-\right)\right)$ to solve the $\mathcal{S}_{(1)}$ equation we inadvertently incorporated a portion of the homogeneous solution corresponding to one of the anistropic degrees of freedom into our solution $\left(39\right)$. As a matter of fact the closed form solution $\left(40\right)$ we obtained can also be found by using Bae's globally defined $\mathcal{S}_{(1)}$ $\left(34\right)$, one of the conserved quantities that he computed 

\begin{equation}
\phi_{(0)} :=\frac{1}{2 \sqrt{3}} e^{4 \alpha-2 \beta_{+}} \sinh \left(2 \sqrt{3} \beta_{-}\right),
\end{equation}
and setting the ordering parameter in $\left(34\right)$ to $ 2\sqrt{33}$. The reader can easily verify the above using the semi-classical method previously described. Finding our closed form solution through only the 'ground' state transport equations as opposed to the $\phi$ equations allows us to easily verify whether the infinite sequence of transport equations terminates by checking to see if our $\mathcal{S}_{(1)}$ causes the source term of the $\mathcal{S}_{(2)}$ equation to vanish without having to switch to a different set of transport equations. 

If the potential to our Bianchi A model can be written as $e^{4\alpha}U\left(\beta_+,\beta_-\right)$, than the following ansatz proposed by Moncrief\cite{moncrief2014euclidean} can be used in principle to solve the $k \geq 2$ 'ground' and 'excited' state transport equations

\begin{equation}
\mathcal{S}_{(k)}^{\mathrm{wh}}=6 e^{-2(k-1) \alpha} \Sigma_{(k)}^{\mathrm{wh}}\left(\beta_{+}, \beta_{-}\right).
\end{equation}

Because of the form of our 'ground' state solutions $\left(18\right)$, using the above ansatz results in a wave function of the following form

\begin{equation}
\stackrel{(0)}{\psi}_{\hbar}=e^{-\frac{1}{X} \mathcal{S}_{(0)}-\mathcal{S}_{(1)}-\left(\sum_{k=2}^{\infty}\frac{X^{k-1}}{k!} 6 e^{-2(k-1) \alpha} \Sigma_{(k)}^{\mathrm{wh}}\left(\beta_{+}, \beta_{-}\right)\right) }
\end{equation}

where it is assumed $\Sigma_{(k)}^{\mathrm{wh}}\left(\beta_{+}, \beta_{-}\right)$ can always be found through the 'ground' state transport equations and that the resultant wave function is always smooth and globally defined. It makes sense physically that the higher order quantum corrections decay exponentially as the size of the universe increases, in essence they get washed  out. However as we will elaborate on later, some quantum corrections don't ever completely wash out no matter how large the  universe becomes, they always exert a noticeable affect on its evolution. Furthermore one does not have to restrict themselves to (45); but rather seek more exotic solutions to the 'ground' state transport equations which yield smooth and globally defined wave functions. 

To compute solutions to the higher order 'excited' state transport equations we first have to establish what our $\phi_{0}$ shall be. For Bianchi A models with two anistropic degrees of freedom their $\phi_{0}$ homogeneous transport equations have the following solutions $C\left(f\left(\alpha,\beta_+,\beta_-\right),g\left(\alpha,\beta_+,\beta_-\right)\right)$ where $C$ is any arbitrary function of $f\left(\alpha,\beta_+,\beta_-\right)$ and $g\left(\alpha,\beta_+,\beta_-\right)$. This set of solutions composed of $f\left(\alpha,\beta_+,\beta_-\right)$ and $g\left(\alpha,\beta_+,\beta_-\right)$ follows from the standard form that solutions to the homogeneous transport equation of three variables have. Both $f\left(\alpha,\beta_+,\beta_-\right)$ and $g\left(\alpha,\beta_+,\beta_-\right)$ are solutions to the $\phi_{0}$ transport equation in their own right, and are independent conserved quantities along the $S_{0}$ flow. Taking advantage of the fact that any function of a conserved quantity is a another conserved quantity this solution can be manipulated into the following form 

\begin{equation}
\phi_{0}= S^{m1} C^{m2},
\end{equation}

where the functions $S^{m1}$ and $ C^{m2}$ are some functions of $f\left(\alpha,\beta_+,\beta_-\right)$ and $g\left(\alpha,\beta_+,\beta_-\right)$, depend on the quantum Bianchi A models which are being studied, and are themselves conserved quantities. A requirement of $\phi_{0}$ is that it must be a smooth and globally defined function. Beyond that requirement one can use aesthetics of the resultant wave function as a guide to choose an appropriate $\phi_{0}$, and in addition use as a guide the ease of solving the higher order 'excited' state equations after choosing an initial $\phi_{0}$. The lack of more rigorous requirements indicates that we do not yet possess a precise mathematical definition of 'excited' states in Wheeler DeWitt quantum cosmology. The parameters $\left(m_1, m_2\right)$ can plausibly be interpreted as graviton excitation numbers for the ultra long wavelength gravitational wave modes embodied in the $\left(\beta_+,\beta_-\right)$ anisotropic degrees of freedom \cite{bae2014quantizing}. To illustrate quantization, in the case when both $S$ and $C$ vanish at certain points in minisuperspace, if $\left(m_1, m_2\right)$ are negative, then $\phi_{0}$ possesses a singularity and results in a wave function which is no longer globally defined and smooth. If $\left(m_1, m_2\right)$ are positive non integers then the derivatives of $\phi_{0}$ won't be smooth and globally defined. Thus to keep our 'excited' states globally defined we have to restrict the values of $\left(m_1, m_2\right)$ to positive integers for the case when $S$ and $C$ vanish at certain points in minisuperspace, thus making our 'excited' states discretized in the same manner as the bound excited states of the quantum harmonic oscillator which are labeled by a positive integer $n$.

If the $\phi_{0}$ which was picked has the form $e^{\left(m1+m2\right)\alpha}Z\left(\beta_+,\beta_-\right)$ where Z is some function of the betas, then to solve the higher order $\phi_{k}$ equations we can use the following ansatz 

\begin{equation}
{\phi}(k)=e^{(4|m|-2 k) \alpha} {\chi}_{(k)}\left(\beta_{+}, \beta_{-}\right)
\end{equation}

where $|m| :=m_{1}+m_{2}$. 

This induces the following form for our 'excited' state wave function. 

\begin{equation}
{\psi}_{\hbar}=\left(\phi_{0}+\sum_{k=1}^{\infty}\frac{X^{k}}{k!}e^{(4|m|-2 k) \alpha} {\chi}_{(k)}\left(\beta_{+}, \beta_{-}\right)\right) e^{-S_{\hbar} / \hbar},
\end{equation}

In addition it simplifies the higher order 'exicted' state transport equations. If one is only interested in the semi-classical limit, and not in explicit higher order quantum corrections they can choose a more exotic $\phi_{0}$. 

A method for proving the existence of smooth globally defined solutions to both the 'ground' and 'excited' state transport equations for the vacuum Bianchi IX model can be found in \cite{moncrief2014euclidean} section 4 and 5 respectively. 

Before we move on to the Bianchi IX wave functions which are restricted to the $\beta_+$ axis we will form a superposition of wave functions using the leading order 'excited' states that Bae and Moncrief \cite{bae2015mixmaster} found and plot them. Joseph Bae found the following two independent solutions to $\phi_{0}$

\begin{equation}
C_{(0)} :=\frac{1}{6} e^{4 \alpha-2 \beta_{+}}\left(e^{6 \beta_{+}}-\cosh \left(2 \sqrt{3} \beta_{-}\right)\right)
\end{equation}

and as previously mentioned $\left(44\right)$.

Using Bae's two conserved quantities we can form the following family of leading order 'excited' states
\begin{equation}
\Psi_{m1,m2}\left(\alpha,\beta_+,\beta_-\right) =e^{\frac{1}{2} \alpha (\text{B}+6)- \mathcal{S}_{(0)}} S^{m1}_{0} C^{m2}_{0}.
\end{equation}

Before we construct our superposition of states we will perform the following averaging over the group associated with the rotational symmetry in the $\left(\beta_+,\beta_-\right)$ plane 
\begin{equation}
\begin{aligned}
\Psi1_{m1,m2} = \Psi_{m1,m2}\left(\alpha,\beta_+,\beta_-\right)
\end{aligned}    
\end{equation}
\begin{equation}
\begin{aligned}
\Psi2_{m1,m2} = \Psi_{m1,m2}\left(\alpha,-\frac{\sqrt{3}\beta_-}{2}-\frac{\beta_+}{2},-\frac{\beta_-}{2}+\frac{\sqrt{3}\beta_+}{2}\right)
\end{aligned}    
\end{equation}
\begin{equation}
\begin{aligned}
\Psi3_{m1,m2} = \Psi_{m1,m2}\left(\alpha,\frac{\sqrt{3}\beta_-}{2}-\frac{\beta_+}{2},-\frac{\beta_-}{2}-\frac{\sqrt{3}\beta_+}{2}\right)
\end{aligned}    
\end{equation}

which leads to our 'excited' states possessing the $\frac{2\pi}{3}$ rotational symmetry of the Bianch IX potential.

\begin{equation}
\begin{aligned}
\Psi_{m1,m2,avg} = \frac{1}{3}\left(\Psi1_{m1,m2}+\Psi2_{m1,m2}+\Psi3_{m1,m2}\right).
\end{aligned}    
\end{equation}

A superposition of states can take the form presented below.

\begin{equation}
\begin{aligned}
\Psi_{superposition} =  \sum_{m1=0}^{\infty} \sum_{m2=0}^{\infty} e^{-m1^{2}}e^{-m2^{2}}\Psi_{m1,m2,avg}
\end{aligned}    
\end{equation} 

Two plots of a superposition of states with m1 and m2 ranging from 0 to 10 are shown in figure 3 and 4.

\begin{figure}[!ht]
\begin{minipage}[c]{0.4\linewidth}
\includegraphics[scale=.1]{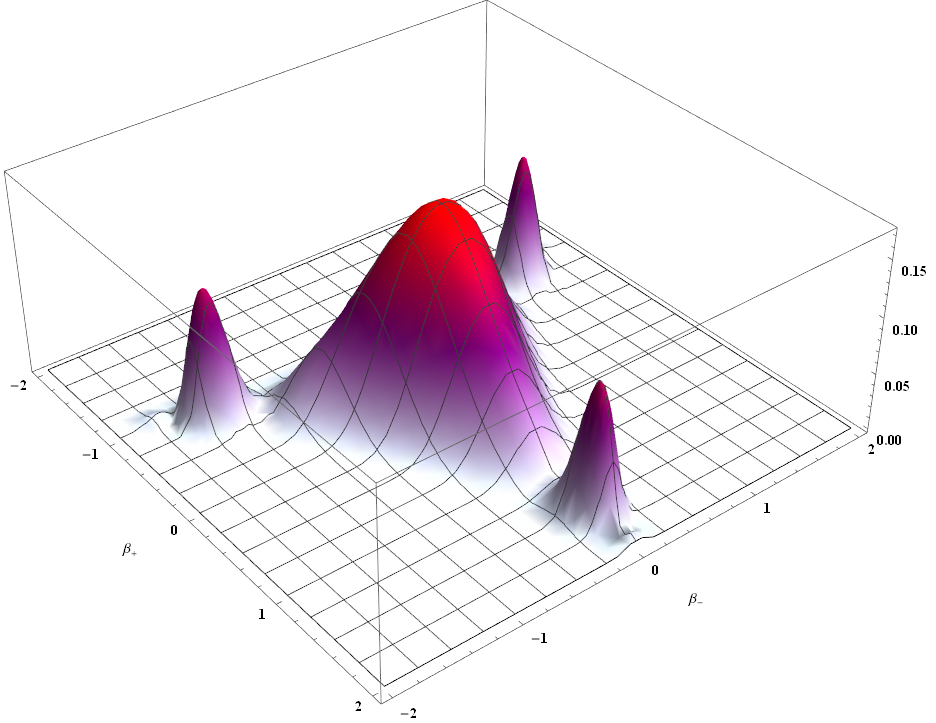}
\caption{$\alpha=0$}
\end{minipage}
\hfill
\begin{minipage}[c]{0.4\linewidth}
\includegraphics[scale=.1]{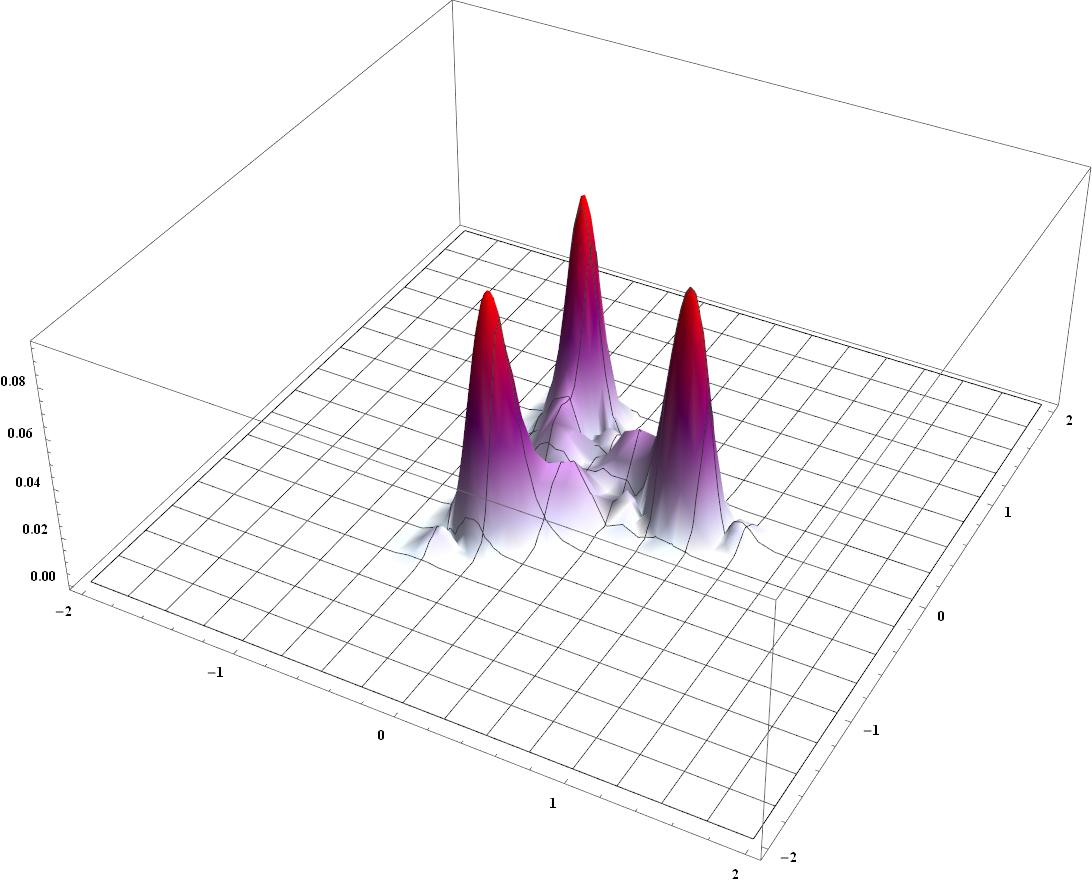}
\caption{$\alpha=1$}
\end{minipage}%
\end{figure}

It should be noted that instead of generating a real valued wave function we could have constructed a complex wave function which would have facilitated the construction of a non trivial Klein-Gordon current.

Unlike our closed form solutions we are able to use group averaging to produce a non trivial leading order superpositon of states. Despite our closed form solutions failing to satisfy the symmetry of the Bianchi IX potential the author is of the belief as of the writing of this manuscript that they should not be considered any less valuable despite this potential shortcoming. If we examine the solutions to the Bianchi IX Euclidean-signature Hamilton Jacobi equation we already see symmetry breaking in the "arm" solutions which are not invariant under a $\frac{2\pi}{3}$ rotation in the $\beta$ plane. Wave functions which are not group averaged which are constructed from these "arm" solutions violate the symmetry of the Bianchi IX potential. Additionally one can forgo the group averaging procedure we applied to the leading order 'wormhole excited' Bianchi IX states, and construct\cite{moncrief2014euclidean} smooth globally defined solutions to the higher order 'excited' state transport equations, which would result in a symmetry breaking asymptotic solution to the Bianchi IX Wheeler Dewitt equation. Furthermore if we were to accept the view that solutions which do not respect the rotational symmetry of the Bianchi IX potential are not physical, we would have to conclude that the sum of three non-physical solutions admits a physical one under group averaging. This is a potentially troublesome view to have. 

Precedence for symmetry breaking in Wheeler DeWitt quantum cosmology can potentially be present in quantum field theory and classical mechanics. Within quantum field theory there exists certain circumstances where the symmetry of the Hamiltonian is not respected by its ground state. When this occurs it is called spontaneous symmetry breaking\cite{van2007spontaneous}, and it is even present in classical physics\cite{ochoa2006bead}. Though in ordinary finite dimensional quantum mechanics spontaneous symmetry breaking is forbidden, a more in depth discussion of this can be found in \cite{landsman2013spontaneous}; however this does not necessarily apply to finite dimensional Wheeler DeWitt quantum cosmology because as we explained earlier it possesses fundamental mathematical differences from ordinary quantum mechanics. Thus we cannot throw out the possibility that spontaneous symmetry breaking occurs in Wheeler DeWitt quantum cosmology. Though it should be noted that the author does not claim that our closed form solutions to the Wheeler DeWitt equation which violate its discrete rotational symmetry in the $\beta$ plane is an example of spontaneous symmetry breaking, nor are the asymptotic solutions that one can construct for the 'excited' states. Nonetheless because of the known precedents in quantum field theory and classical mechanics for the symmetries of the equations of motion to be violated, and our current incomplete understanding of quantum gravity/cosmology, despite our solutions not respecting the symmetry of the classical Bianchi IX potential, in this paper we will hold them in the same regards as we do the $B= \pm 6$ solutions which do respect the symmetry of the potential. The author hopes that the above short discussion spurs further discussion in the physics community about the nature of symmetry breaking in Wheeler DeWitt quantum cosmology.

\section{\label{sec:level1} Bianchi IX wave functions restricted to the $\beta_+$ axis}

Before we can construct wave functions which are restricted to the $\beta_+$ axis we must show that classically the 'no boundary' solution $\left(30\right)$ and the "arm3" solution $\left(33\right)$ admit non trivial flows in minisuperspace along the $\beta_+$ axis. In addition it should be stressed that what we are doing is not equivalent to finding Wheeler DeWitt wave functions for the quantum Bianchi IX LRS models. Even though we will be setting $\beta_-=0$ for some parts of our calculation, as will be seen shortly the sheer existence of a $\beta_-$ axis will affect how we proceed. The author has already compiled a large number of preliminary results for the quantum  Bianchi IX LRS model using this Eucldiean-signature semi classical method and plans to publish them in a future work. For the remainder of this section we will only consider the + sign form of these solutions to the Euclidean-signature Hamiliton-Jacobi equation. Furthermore we will work under the assumption that smooth globally defined solutions do exist for the higher order transport equations for both the 'no boundary' and "arm3" case. Using the following equations
\begin{equation}
\begin{aligned} 
p_{\alpha} = \frac{\partial \mathcal{S}_{(0)}}{\partial \alpha} \\ p_{+} = \frac{\partial \mathcal{S}_{(0)}}{\partial \beta_{+}} \\ p_{-} = \frac{\partial \mathcal{S}_{(0)}}{\partial \beta_{-}}
\end{aligned}
\end{equation}

\begin{equation}
\begin{aligned} \dot{\alpha} &=\left.\frac{(6 \pi)^{1 / 2}}{2 e^{3 \alpha}} N\right|_{\text {Eucl }} p_{\alpha} \\ \dot{\beta}_{+} &=\left.\frac{-(6 \pi)^{1 / 2}}{2e^{3 \alpha}} N\right|_{\text {Eucl }} p_{+} \\ \dot{\beta}_{-} &=\left.\frac{-(6 \pi)^{1 / 2}}{2e^{3 \alpha}} N\right|_{\text {Eucl }} p_{-} \end{aligned}
\end{equation}

where $ N|_{\text {Eucl }}$ is the lapse, we can construct the flow equations and easily show that classically flows from the two aforementioned solutions to the Hamilton Jacobi equation that start on the $\beta_+$ axis remain on the $\beta_+$ axis. It should be mentioned that the $Eucl$ part means that these are the flow equations for the minisuperspace variables of the Euclidean as opposed to Lorentzian signature Bianchi IX models, though this won't affect our conclusion. The lapse $ N|_{\text {Eucl }}$ can be any function of the Misner variables as long as it never vanishes or changes sign within the range $-\infty$ to $\infty$ of all three variables. To keep things simple though we will set $ N|_{\text {Eucl }}=\frac{2 e^{a}}{\left(6\pi\right)^{1/2}}$ and first write out the flow equations $\left(58\right)$ for the 'no boundary' case 

\begin{equation}
\begin{aligned}
\frac{d \alpha}{d t}=\frac{1}{3} e^{-4 \beta_+} \Bigl(4 e^{6 \beta_+} \sinh ^2\left(\sqrt{3} \beta_-\right) \\-4 e^{3 \beta_+} \cosh
   \left(\sqrt{3} \beta_-\right)+1\Bigr)
 \end{aligned} 
\end{equation}

\begin{equation}
\begin{aligned}
\frac{d \beta_{+}}{d t}=-\frac{2}{3} e^{-4 \beta_+}  \Bigl(2 e^{6 \beta_+} \sinh ^2\left(\sqrt{3} \beta_-\right) \\ +e^{3 \beta_+} \cosh
   \left(\sqrt{3} \beta_-\right)-1\Bigr)
\end{aligned}
\end{equation}

\begin{equation}
\begin{aligned}
\frac{d \beta_{-}}{d t}=-\frac{2 e^{-\beta_+} \sinh \left(\sqrt{3} \beta_-\right) \left(2 e^{3 \beta_+} \cosh \left(\sqrt{3}
   \beta_-\right)-1\right)}{\sqrt{3}}.
\end{aligned}
\end{equation}

Because $\frac{d \beta_{-}}{d t}$ vanishes when $\beta_-=0$ if the flow dictated by the 'no boundary' Hamiliton Jacobi solution in minisuperspae were to start on the $\beta_+$ axis it would remain on that axis for all "time". Therefore it makes sense to discuss the quantum analogue of this which would be a Wheeler Dewitt wave function which vanishes $\beta_-$ axis, and is peaked on the $\beta_+$ axis. For the "arm3" solution we obtain the following flow equations

\begin{equation}
\begin{aligned}
\frac{d \alpha}{d t}=\frac{1}{3} e^{-4 \beta_+} \Bigl(4 e^{6 \beta_+} \sinh ^2\left(\sqrt{3} \beta_-\right) \\ +4 e^{3 \beta_+} \cosh
   \left(\sqrt{3} \beta_-\right)+1 \Bigr)
 \end{aligned} 
\end{equation}

\begin{equation}
\begin{aligned}
\frac{d \beta_{+}}{d t}=-\frac{2}{3} e^{-4 \beta_+} \Bigl( e^{6 \beta_+} \sinh ^2\left(\sqrt{3} \beta_-\right) \\ -e^{3 B} \cosh
   \left(\sqrt{3} \beta_-\right)-1\Bigr)
\end{aligned}
\end{equation}

\begin{equation}
\begin{aligned}
\frac{d \beta_{-}}{d t}=-\frac{2 e^{-\beta_+} \left(e^{3 \beta_+} \sinh \left(2 \sqrt{3} \beta_-\right)+\sinh \left(\sqrt{3}
   \beta_-\right)\right)}{\sqrt{3}}.
\end{aligned}
\end{equation}

Just like the 'no boundary' case, $\frac{d \beta_{-}}{d t}$ vanishes when $\beta_-=0$, thus a flow in minisuperspace caused by this "arm" solution which starts on the $\beta_+$ axis will stay on the $\beta_+$ axis. The "arm3" solution that we choose is the only "arm" solution which exhibits this property, as a result the classical cosmology of the "arm3" solution restricted to the $\beta_+$ axis has a quantum counterpart. 
We will first compute an approximate wave function restricted to the $\beta_+$ axis for the 'no boundary' proposal. To begin we will substitute $\left(30\right)$ into our $\mathcal{S}_{(1)}$ transport equation, and only after taking all of the required derivatives of $\left(30\right)$ set $\beta_-=0$. This step separates what we are doing from just finding the wave function for the quantum 'no boundary' quantum Bianchi IX LRS model. Because the full Bianchi IX wave function is a function of both of the betas, its derivatives with respect to $\beta_-$ may not vanish at $\beta_-=0$. As a result of inserting $\left(30\right)$ into our $\mathcal{S}_{(1)}$ equation, and then setting $\beta_-=0$ we obtain the following transport equation.  

\begin{equation}
\begin{aligned}
-4 e^{3 \beta_+} (2 \frac{\partial \mathcal{S}_{(1)}}{\partial \alpha}+\frac{\partial \mathcal{S}_{(1)}}{\partial \beta_+}+\text{B})+2 \frac{\partial \mathcal{S}_{(1)}}{\partial \alpha}+4 \frac{\partial \mathcal{S}_{(1)}}{\partial \beta_+}
\\+12 e^{6 \beta_+}+\text{BB}+6=0
\end{aligned}
\end{equation}

which has this as the following solution

\begin{equation}
\begin{aligned}
\mathcal{S}_{(1)}:=\alpha \left(3-\frac{\text{B}}{2}\right)-3 \beta_++e^{3 \beta_+}.
\end{aligned}
\end{equation}

Using the $\mathcal{S}_{(1)}$ that we calculated we can construct a first order solution to the Bianchi IX Wheeler DeWitt equation which is valid when $\beta_-=0$. Unfortunately though we cannot go beyond first order with this method. If we wanted to go beyond the first order quantum correction we would need to know what the first and second derivatives of the smooth and globally defined function which satisfies the full $\mathcal{S}_{(1)}$ transport equation is with respect to $\beta_-$. Because of how formidable the full 'no boundary' $\mathcal{S}_{(1)}$ transport equation is, obtaining a closed form solution to it is not trivial and as a result that is why in this work we are only considering a solution for a restricted case. Even though we cannot go beyond the first order transport equation, under the assumptions we outlined at the beginning of this section, and assuming our wave function takes the form of the aforementioned ans$\text{\" a}$tze these approximate solutions become increasingly accurate as $\alpha$ grows larger because solutions to the higher order transport equations decays exponentially as $\alpha$ increases. Thus our wave functions within the context of our assumptions are useful for obtaining a qualitative picture of how the 'no boundary' wave function and the wave function associated with the "arm3" solution $\left(33)\right)$ behave when $\alpha>>0$, when first order quantum affects are considered. They can also be used to explore purely the semi-classical limit. 

In the same way we solved the $\mathcal{S}_{(1)}$ transport equation for the 'no boundary' solution by restricting ourselves to the $\beta_+$ axis we can solve the $\phi_{0}$ equation and construct leading order 'no boundary' 'excited' states. Doing the same calculation as above except substituting the 'no boundary' solution into the $\phi_{0}$ equation results in this conserved quantity which can be used to construct a leading order 'excited' state 

\begin{equation}
\phi_{0 \hspace{1mm} nb}:=\left(\left(1-e^{3 \beta_+}\right) e^{\beta_+-2 \alpha}\right)^{\text{m1}}
\end{equation}

where $m1$ is a positive integer. Using our results we can construct the following approximate wave function restricted to the $\beta_+$ axis for the 'no boundary' case, which becomes increasingly accurate as $\alpha$ grows
\begin{equation}
\psi_{nb}=\phi_{0 \hspace{1mm} nb} e^{\frac{1}{2} a (\text{B}-6)+3 \beta_+-e^{3 \beta_+} \mp \mathcal{S}_{(0 \hspace{1mm} nb)}^{\beta_-=0}}
\end{equation}
two different plots of its superposition of 'ground and 'excited' states are displayed below.

\begin{figure}[!ht]
\begin{minipage}[c]{0.4\linewidth}
\includegraphics[scale=.1]{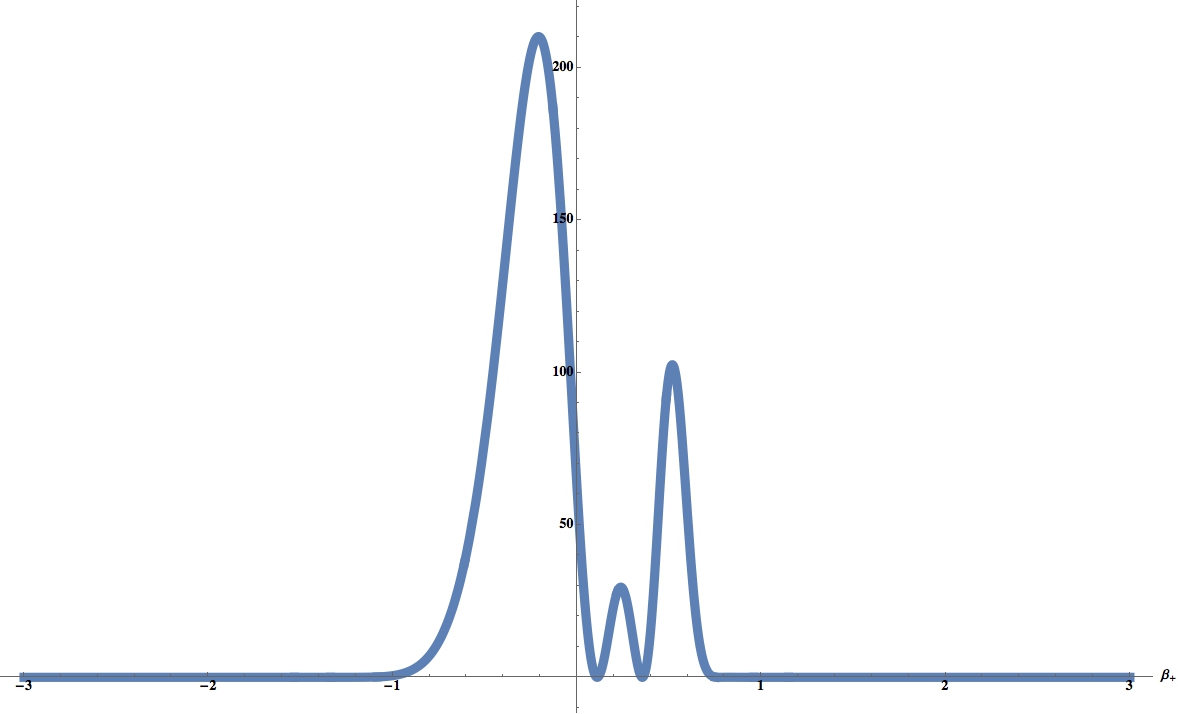}
\caption{$ \alpha=-1$ \hspace{1 mm} \\ $| \sum_{m1=0}^{10} e^{-m1^{2}}\psi_{nb}|$}
\end{minipage}
\hfill
\begin{minipage}[c]{0.4\linewidth}
\includegraphics[scale=.101]{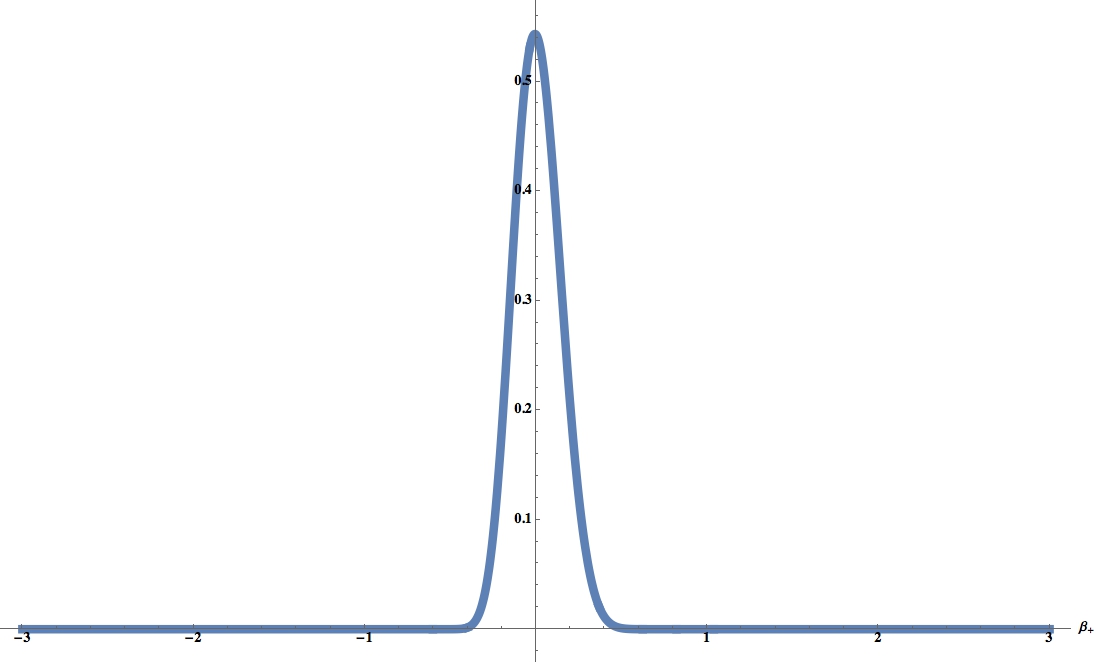}
\caption{$ \alpha=2$ \hspace{1 mm} \\ $| \sum_{m1=0}^{10} e^{-m1^{2}}\psi_{nb}|$}
\end{minipage}%
\end{figure}

We included a graph for $\alpha=-1$ which depicts some interesting semi-classical behavior. 

Before applying the same techniques that we brought to bare on the 'no boundary' case we should mention that our "arm3" solution when $\beta_-=0$ possesses some pathologies. As the reader can verify the wave function $e^{\mp \mathcal{S}_{(0 arm3)}}$ when $\beta_-=0$ does not behave as a traveling Gaussian like wave for any value of $\alpha$. In addition as will be shown below our $\mathcal{S}_{(1)}$ has a pathological term in it as well. To somewhat remedy this and obtain a traveling wave which is amenable to a straight forward interpretation, we will choose a particular $\phi_{0}$ which has the effect of offsetting the behavior caused by the aforementioned pathological terms.

As it can be seen below our $\mathcal{S}_{(1)}$ contains the term $-e^{3\beta_+}$ which on its own is pathological because when it is inserted into the form of our wave function $\left(18\right)$, it results in a wave function which looks nothing like a Gaussian and increases without bound as $\beta_+$ grows larger. 

\begin{equation}
\begin{aligned}
\mathcal{S}_{(1)}:=-\frac{1}{4} \beta_+ (\text{B}+6)-\frac{1}{4} (\text{B}-6) \log \left(e^{3 \beta_+}+1\right)-e^{3 \beta_+}
\end{aligned}
\end{equation}

To remedy this we will take advantage of the fact that any function of a conserved quantity is itself a conserved quantity. In doing so we will choose the following to be our $\phi_{0}$

\begin{equation}
\begin{aligned}
\phi_{0 \hspace{1 mm} arm3}:=e^{\left(e^{3 \beta_+}+1\right) \left(-e^{\beta_+-2 \alpha}\right)}  \left(\left(e^{3 \beta_+}+1\right)e^{\beta_+-2 \alpha}\right)^{m1}
\end{aligned}
\end{equation}

where $m_{1}$ can be any real or complex number. This $\phi_{0}$ remedies the pathological behavior which results from $\left(69\right)$ by including in the exponent of our wave function a term which contains a  $e^{4\beta_+}$ that has the opposite sign of the one in $\left(69\right)$. Combining what we computed above results in the following "arm3" $\beta_+$ axis wave function:

\begin{equation}
\psi_{arm3  \hspace{1 mm}}= \phi_{0 \hspace{1 mm} arm3} e^{\frac{1}{2} \alpha (\text{B}-6)+3 \beta_+ +e^{3 \beta_+} \mp \mathcal{S}_{(0 \hspace{1mm} arm3)}^{\beta_-=0}}.
\end{equation}

Even though our $\beta_+$ axis "arm3" wave functions have the form of 'excited' states, qualitatively, as the plot below shows, and as the reader can further verify they aesthetically behave like 'ground' states. This is perhaps because we chose an unusual form for our $\phi_{0}$ which in principle complicates the solving of the higher order $\phi_{k}$ transport equations. Another possibility is that 'excited' states do not exist for the "arm3" solution restricted to the $\beta_+$ axis. Because we don't possess in Wheeler DeWitt quantum cosmology yet a rigorous understanding of what constitutes a 'ground' vs 'excited' state like we do in ordinary quantum mechanics, and as this exercise has shown we have to be cautious in how we denote states. Just because we use solutions to the aforementioned 'ground' or 'excited' state transport equations doesn't necessarily mean that wave functions constructed from those solutions are ground or excited states, and hence as mentioned before at this point it is best to appeal to their qualitative features to make that determination. 

\begin{figure}[!ht]
\centering
\includegraphics[scale=.1]{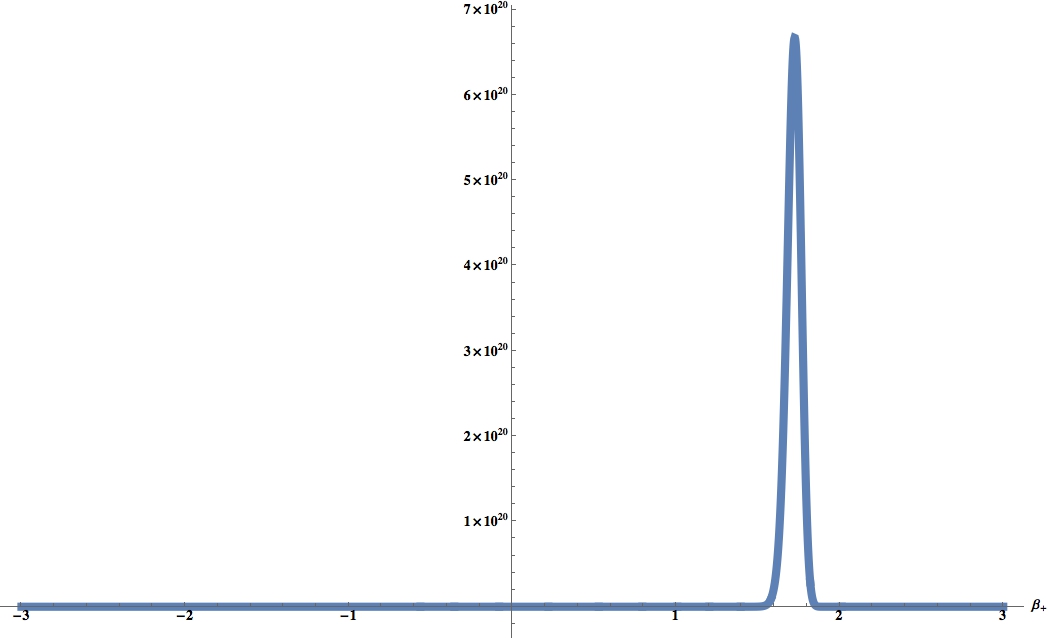}
\caption{For $\alpha=1$  \hspace{1 mm} $\psi_{arm3  \hspace{1 mm} }$} 
\label{fig:my_label}
\end{figure}

Returning to wave function in the full $\beta$ plane, if we assume globally defined solutions for arbitrary Hartle Hawking ordering parameters exist for the transport equations associated with the 'no boundary', and arm solutions as was proven for the 'wormhole' case then it is instructive to take superpositions of our leading order states constructed from $\left(29-33\right)$. Because the "arm" solutions are related to each other by $\frac{2\pi}{3}$ rotations in the $\beta$ plane in principal one only needs to show that globally defined solutions to the higher order transport equations exist for one of them to prove that they exist for the other two. Below we will show six different plots. The first two are group averaged "arm" solutions for two different values of $\alpha$. As it is clear from observing the first two plots the wave functions corresponding to the "arm" solutions possess very different characteristics than the wave functions of the 'wormhole' solutions. The next two plots are a superposition of both the "arm" wave function and our 'excited' state and 'ground' state 'wormhole' wave functions. The last two are a superposition constructed from the "arm" solutions, the 'no boundary' solution and the 'wormhole' solutions. We will discuss what these amazing looking graphs mean towards the end of this work. 

\begin{figure}[!ht]
\begin{minipage}[c]{0.4\linewidth}
\includegraphics[scale=.1]{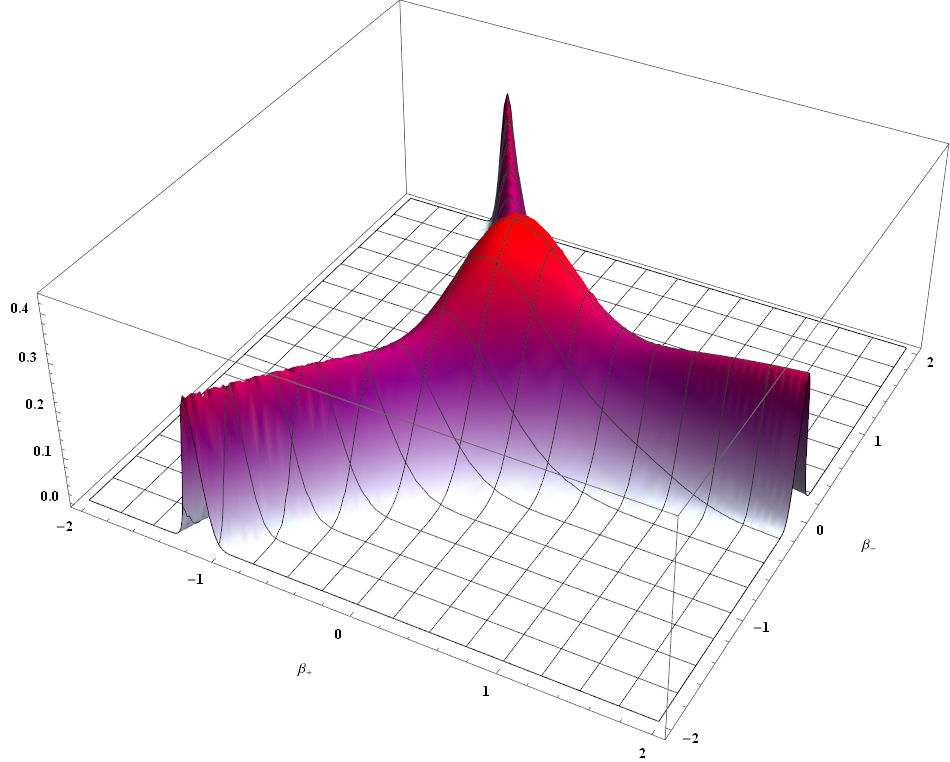}
\caption{$\alpha=0$}
\end{minipage}
\hfill
\begin{minipage}[c]{0.4\linewidth}
\includegraphics[scale=.1]{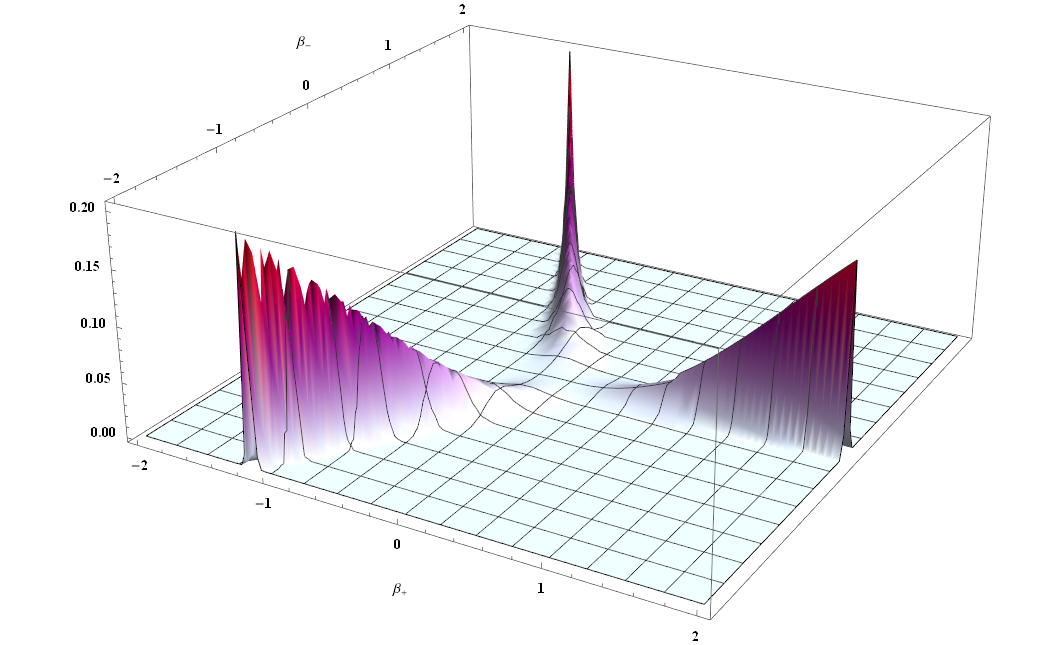}
\caption{$\alpha=1$}
\end{minipage}%
\end{figure}

\begin{figure}[!ht]
\begin{minipage}[c]{0.4\linewidth}
\includegraphics[scale=.1]{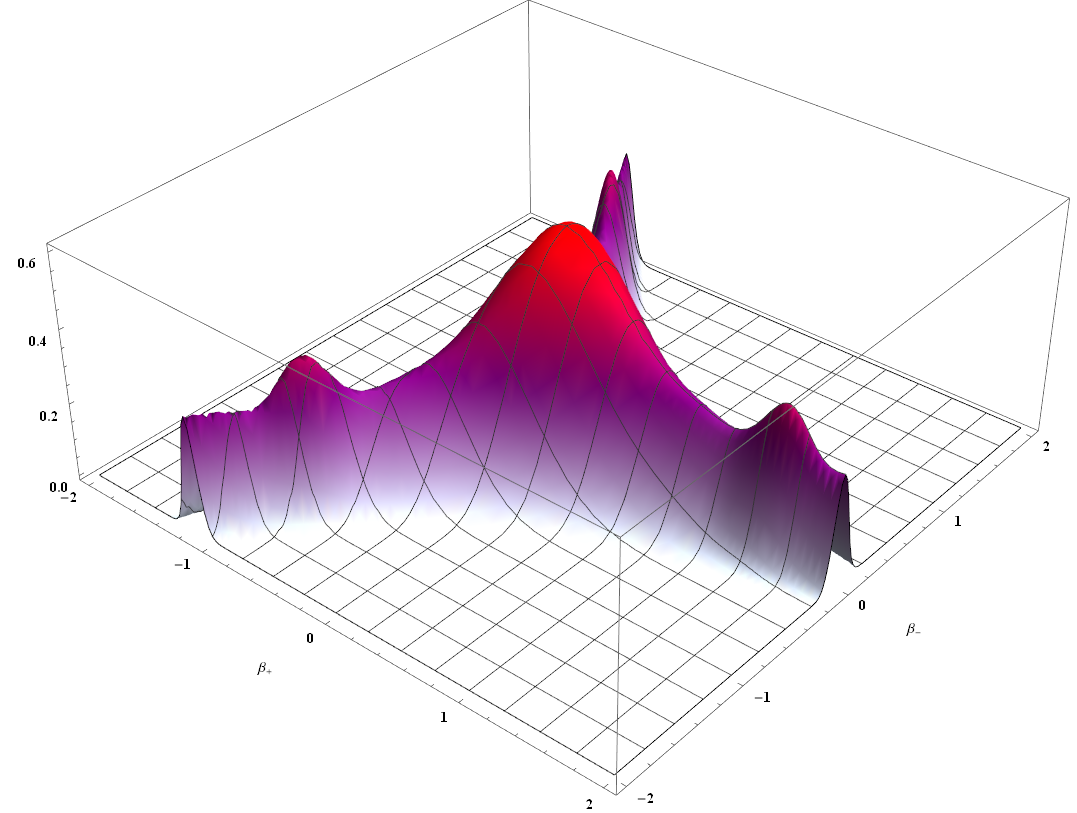}
\caption{$\alpha=0$}
\end{minipage}
\hfill
\begin{minipage}[c]{0.4\linewidth}
\includegraphics[scale=.1]{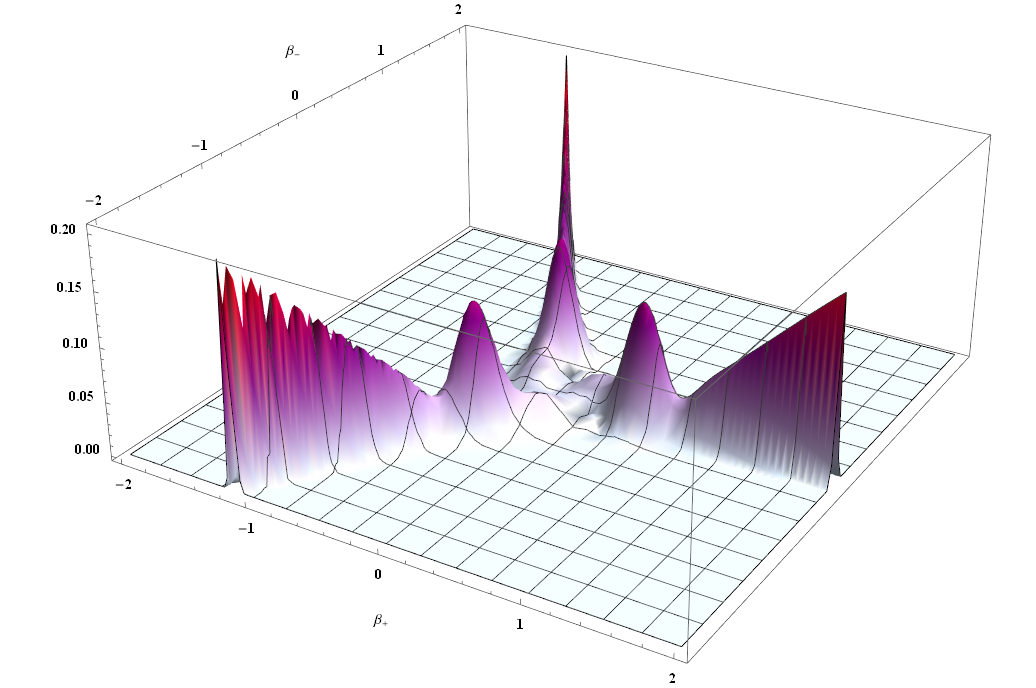}
\caption{$\alpha=1$}
\end{minipage}%
\end{figure}

\begin{figure}[!ht]
\begin{minipage}[c]{0.4\linewidth}
\includegraphics[scale=.1]{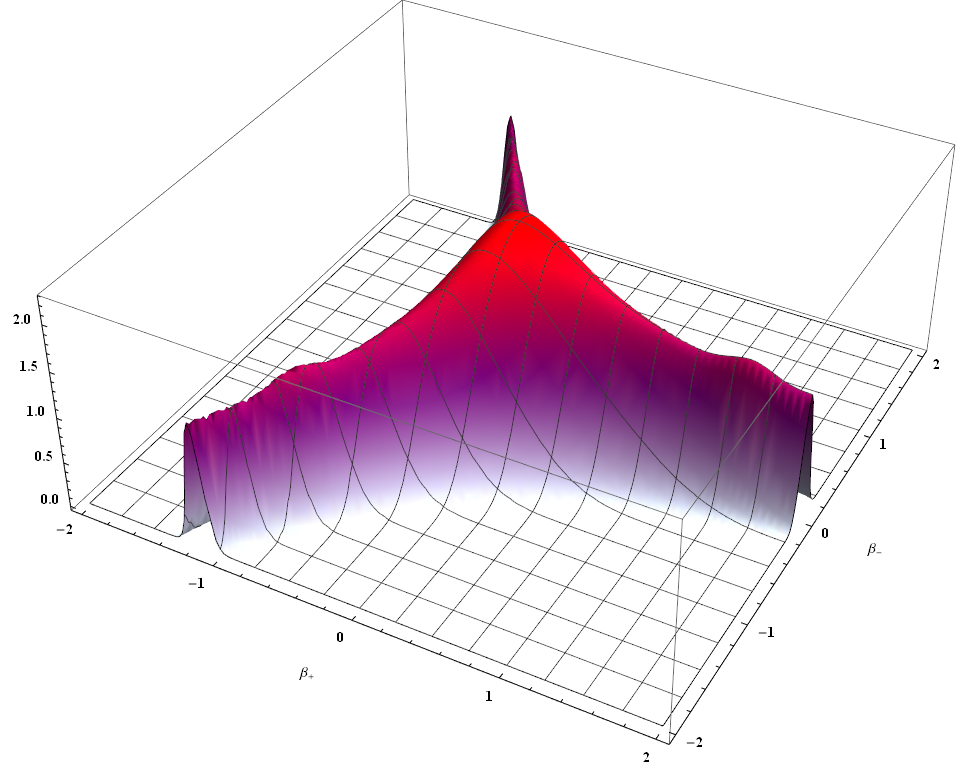}
\caption{$\alpha=0$}
\end{minipage}
\hfill
\begin{minipage}[c]{0.4\linewidth}
\includegraphics[scale=.1]{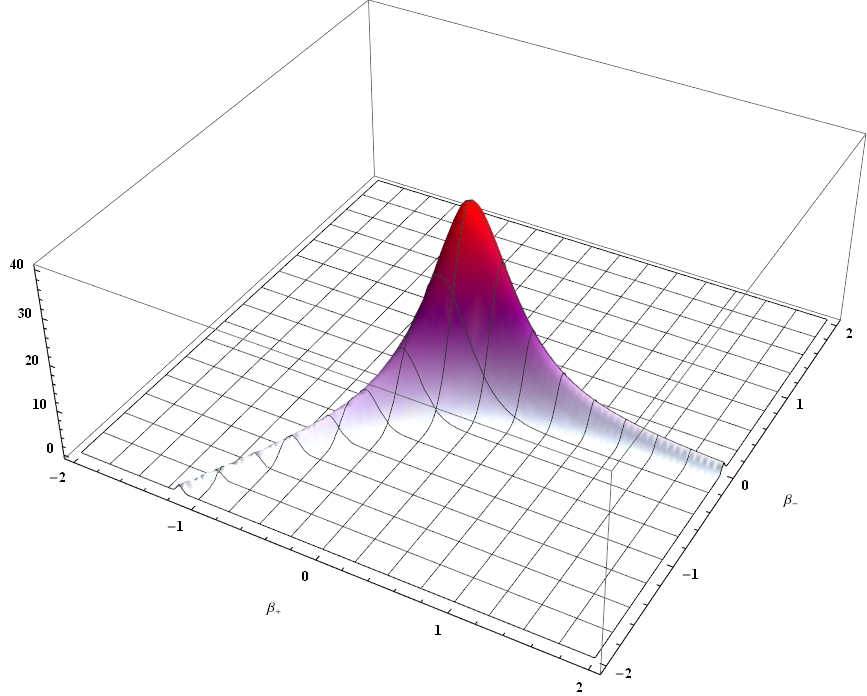}
\caption{$\alpha=1$}
\end{minipage}%
\end{figure}

\section{\label{sec:level1}Leading Order Solutions For The Bianchi IX Symmetry Reduced Wheeler DeWitt Equation With Cosmological Constant and Primordial Magnetic Field}

To obtain a leading order solution for $\Lambda$ $\neq$ 0  we seek  $S_{(0)}$ solutions to the following Euclidean-signature Hamilton Jacobi equation 

\begin{equation}
\begin{aligned}
&{\left(\frac{\partial \mathcal{S}_{(0)}}{\partial \alpha}\right)^{2}-\left(\frac{\partial \mathcal{S}_{(0)}}{\partial \beta_{+}}\right)^{2}-\left(\frac{\partial \mathcal{S}_{(0)}}{\partial \beta_{-}}\right)^{2}}+V=0\\&
V=e^{4\alpha}\Biggl[\frac{e^{-8\beta_+}}{3}-\frac{4}{3}e^{-2\beta_+}\cosh{\left(2\sqrt{3}\beta_-\right)} \\ & + \frac{2}{3}e^{4\beta_+}\left(\cosh{\left(4\sqrt{3}\beta_-\right)}-1\right)+\frac{2e^{2\alpha} \Lambda}{9\pi}\Biggr]
\end{aligned}
\end{equation}

The author has found six solutions to this equation which are the following
\begin{equation}
\begin{aligned}
\mathcal{S}^1_{(\pm 0)}:= & \pm \Biggr( \frac{1}{6} e^{2 \alpha} \left(2 e^{2 \beta_+} \cosh \left(2 \sqrt{3} \beta_-\right)+e^{-4
   \beta_+}\right)-\\ &\frac{\Lambda e^{4 \alpha+4 \beta_+}}{36 \pi } \Biggl)
\end{aligned}
\end{equation}
\begin{equation}
\begin{aligned}
\mathcal{S}^2_{\pm (0)}:= \pm & \Biggr( \frac{1}{6} e^{2 \alpha} \left(2 e^{2 \beta_+} \cosh \left(2 \sqrt{3} \beta_-\right)+e^{-4
   \beta_+}\right)- \\ & \frac{\Lambda e^{4 \alpha-2 \sqrt{3} \beta_--2 \beta_+}}{36 \pi }\Biggl)
   \end{aligned}
\end{equation}
\begin{equation}
\begin{aligned}
\mathcal{S}^3_{\pm (0)}:= \pm &  \Biggr( \frac{1}{6} e^{2 \alpha} \left(2 e^{2 \beta_+} \cosh \left(2 \sqrt{3} \beta_-\right)+e^{-4
   \beta_+}\right)- \\ & \frac{\Lambda e^{4 \alpha+2 \sqrt{3} \beta_--2 \beta_+}}{36 \pi }\Biggl)
   \end{aligned}
\end{equation}

$\mathcal{S}^1_{(\pm 0)}$ is related to $\mathcal{S}^2_{(\pm 0)}$ and $\mathcal{S}^3_{(\pm 0)}$ by a $\frac{2\pi}{3}$ and a $\frac{4\pi}{3}$ rotation in the $\beta$ plane respectively. This allows one to construct leading order non vanishing solutions to the Wheeler DeWitt equation which respect the symmetry of the Bianchi IX potential by group averaging.

\begin{equation}
\psi''= \frac{1}{3}\left(e^{-\mathcal{S}^1_{(+ 0)}}+e^{-\mathcal{S}^2_{(+ 0)}}+e^{-\mathcal{S}^3_{(+ 0)}}\right)
\end{equation}

Like our previous $\mathcal{S}_{(0)}$ solutions, the minus sign solutions are pathological in the sense that as either $\beta_+$ or $\beta_-$ approaches infinity so does the value of the wave functions. However, in addition to the minus sign solutions being pathological so are the solutions for $\Lambda>0$. This does not necessarily mean that the minus sign or $\Lambda>0$ solutions are nonphysical, but rather their physical meaning is obscured. We will not focus on wave functions with those leading order terms in this paper despite including them as solutions to the Euclidean-signature Hamilton Jacobi equation. The plus signed solutions with negative $\Lambda<0$ are well behaved, and thus those are the ones we will consider in this section. 

The author has found the following $\mathcal{S}_{(1)}$ for $\mathcal{S}^1_{(\pm 0)}$ 
\begin{equation}
\begin{aligned}
\mathcal{S}_{(1)}=& \frac{1}{2} \left(- (\text{B}+2)\alpha+\log \left(\sinh \left(2 \sqrt{3} \beta_-\right)\right)-2
   \beta_+\right)
\end{aligned}
\end{equation}

Unfortunately when this solution is exponentiated in the way required by our ansatz it does not yield a globally defined wave function. It is important that our solutions be globally defined so they manifest genuine quantum effects. One can find solutions to the Schr$\text{\" o}$dinger equation for a harmonic oscillator which are not globally defined and do not admit a discrete spectrum. Such states despite satisfying the Schr$\text{\" o}$dinger equation for a harmonic oscillator do not encapsulate its genuine physics. Perhaps in Wheeler Dewitt quantum cosmology genuine quantum effects can be captured in non globally defined solutions, however the author does not have an argument in support of that assertion, and he strongly suspects otherwise. If we work under the assumption that globally defined solutions to the transport equations associated with  $\left(73\right)$ exist to all orders then we can construct a leading order solution which we display graphically in figure 14 and 15. 

\begin{figure}[!ht]
\begin{minipage}[c]{0.4\linewidth}
\includegraphics[scale=.3]{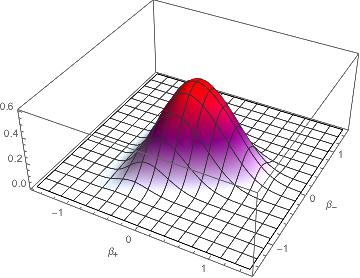}
\caption{$\alpha=0$}
\end{minipage}
\hfill
\begin{minipage}[c]{0.4\linewidth}
\includegraphics[scale=.12]{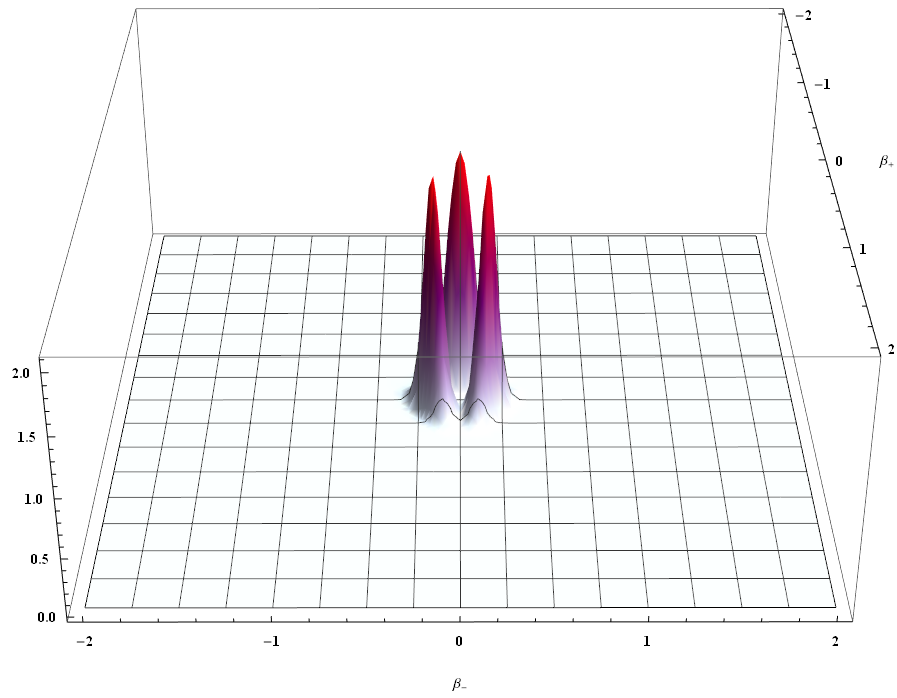}
\caption{$\alpha=2$}
\end{minipage}%
\end{figure}

For the case with a cosmological constant in conjunction with a magnetic field the equation we want to solve is the following, and below it are the solutions the author found

\begin{equation}
\begin{aligned}
&{\left(\frac{\partial \mathcal{S}_{(0)}}{\partial \alpha}\right)^{2}-\left(\frac{\partial \mathcal{S}_{(0)}}{\partial \beta_{+}}\right)^{2}-\left(\frac{\partial \mathcal{S}_{(0)}}{\partial \beta_{-}}\right)^{2}}+V=0\\&
V=e^{4\alpha}\Biggl[\frac{e^{-8\beta_+}}{3}-\frac{4}{3}e^{-2\beta_+}\cosh{\left(2\sqrt{3}\beta_-\right)} \\ & + \frac{2}{3}e^{4\beta_+}\left(\cosh{\left(4\sqrt{3}\beta_-\right)}-1\right)+2b^{2}e^{-2\alpha-4\beta_+}+\frac{2e^{2\alpha} \Lambda}{9\pi}\Biggr]
\end{aligned}
\end{equation}

\begin{equation}
\begin{aligned}
\mathcal{S}^4_{(\pm 0)}:= & \pm \Biggr( \frac{1}{6} \Bigl(2 e^{2 (\alpha+\beta_+)} \cosh \left(2 \sqrt{3} \beta_-\right)-6 \text{b}^2 (\alpha+\beta_+) \\& +e^{2 \alpha-4 \beta_+}\Bigr)-\frac{V e^{4 \alpha+4 \beta_+}}{36 \pi } \Biggl).
\end{aligned}
\end{equation}

The primordial nature of our magnetic field manifests in our solutions by the fact that for $\alpha<<0$ the field portion dominates because it is a linear term and doesn't decay exponentially, while for $\alpha>>0$ it is negligible. The same quantum correction $\left(77)\right)$ that was computed earlier in this section also satisfies the $\mathcal{S}_{(1)}$ equation that follows from $\mathcal{S}^4_{(\pm 0)}$. If we assume that other solutions which are globally defined exist for all of the higher order transport equations for $\mathcal{S}^4_{(\pm 0)}$ then we can construct leading order solutions which become increasingly more accurate as $\alpha$ grows, due to the expected decay of the quantum corrections. Two plots of these leading order solutions are shown below.

\begin{figure}[!ht]
\begin{minipage}[c]{0.4\linewidth}
\includegraphics[scale=.1]{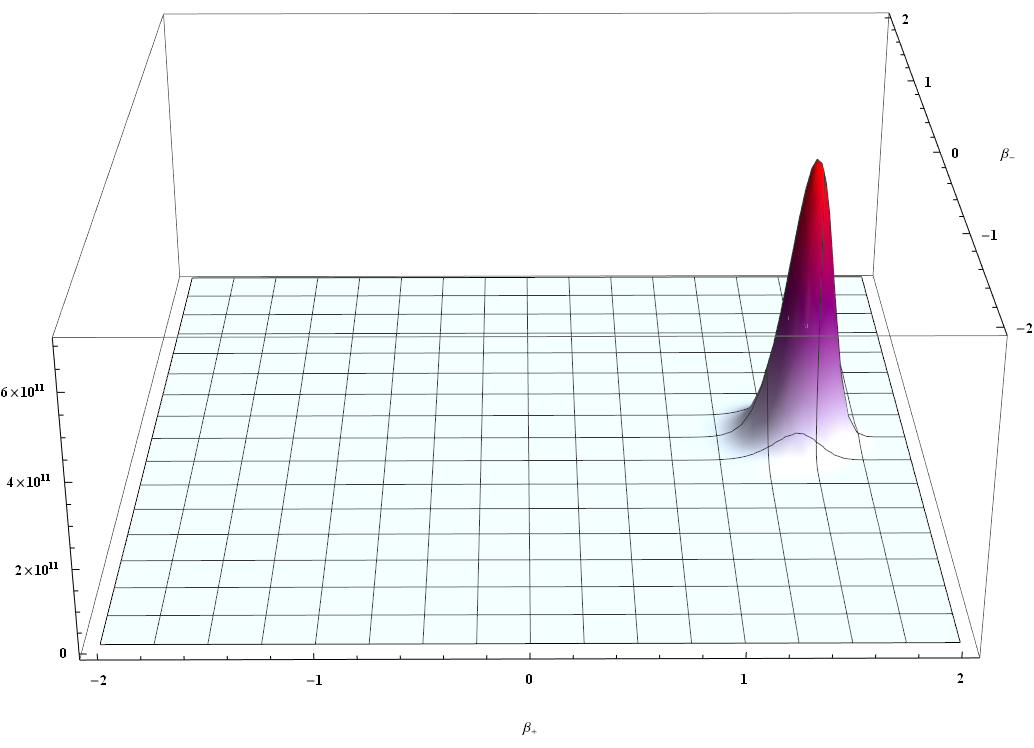}
\caption{ $\alpha=0$ \hspace{1 mm} $\Lambda=-1$ \hspace{1 mm} b=5}
\end{minipage}
\hfill
\begin{minipage}[c]{0.4\linewidth}
\includegraphics[scale=.125]{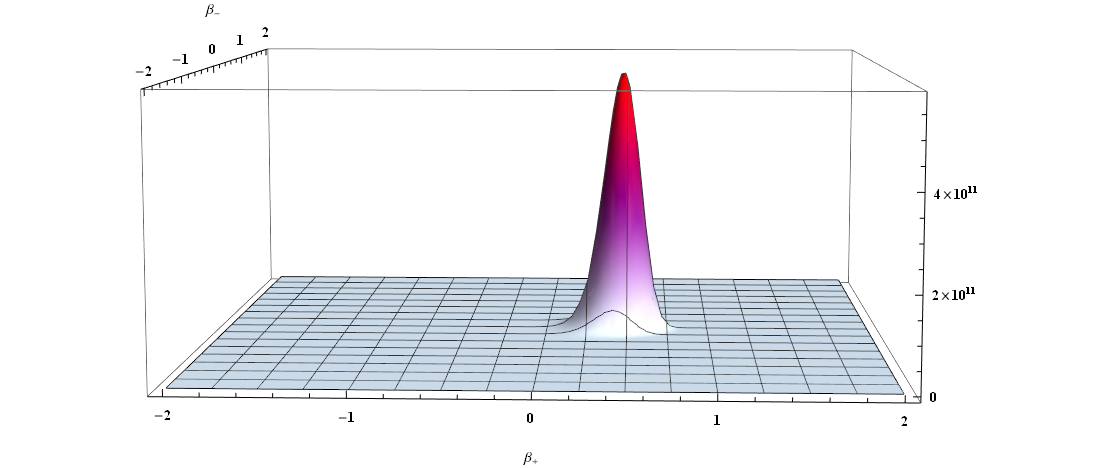}
\caption{ $\alpha=2$ \hspace{1 mm} $\Lambda=-1$ \hspace{1 mm} b=5}
\end{minipage}%
\end{figure}

We can make more progress if we restrict our aforementioned solutions to the $\beta_+$ axis. The flow equations associated with $\left(79\right)$ with the same lapse that we chose before are 

\begin{equation}
\begin{aligned}
\frac{d \alpha}{d t}=\frac{1}{9} \Biggl(-\frac{\Lambda e^{2 \alpha+4 \beta_+}}{\pi }-9 e^{-2 \alpha} \text{b}^2 \\ +6 e^{2 \beta_+} \cosh \left(2 \sqrt{3} \beta_-\right)+3 e^{-4 \beta_+}\Biggr)
 \end{aligned} 
\end{equation}

\begin{equation}
\begin{aligned}
\frac{d \beta_{+}}{d t}=\frac{1}{9} \Biggl(\frac{\Lambda e^{2 \alpha+4 
\beta_+}}{\pi }+9 e^{-2 \alpha} \text{b}^2 \\ -6 e^{2 \beta_+} \cosh \left(2 \sqrt{3} \beta_-\right)+6 e^{-4 \beta_+}\Biggr)
\end{aligned}
\end{equation}

\begin{equation}
\begin{aligned}
\frac{d \beta_{-}}{d t}=-\frac{2 e^{2 \beta_+} \sinh \left(2 \sqrt{3} 
\beta_-\right)}{\sqrt{3}}
\end{aligned}
\end{equation}

and as it can be seen classically a flow that starts on the $\beta_+$ axis remains on that axis. This means that we are justified to construct wave functions like we did in the previous section which are constrained to the $\beta_+$ axis. 

A set of conserved quantities which can be used to construct 'excited' states constrained to the $\beta_+$ axis for our quantum Bianchi IX models with a cosmological constant and primordial magnetic field are

\begin{equation}
\begin{aligned}
\phi_{0 \Lambda}= \Biggl(9 \pi  b^2 e^{2 (\alpha+\beta_+)}+\frac{1}{3} V e^{6 (\alpha+\beta_+)} \\+  3 \pi  e^{4 \alpha-2 \beta_+}-3 \pi  e^{4 (\alpha+\beta_+)}\Biggr)^{m1}
\end{aligned}
\end{equation}

where m1 is a positive integer. Figures 18, 19 and 20 are three plots for a superposition of states with different values for the cosmological constant and the aligned primordial magnetic field. Our states have the following mathematical form $| \sum_{m1=0}^{10} e^{-m1^{2}}\psi_{\Lambda}|$.

\begin{equation}
\begin{aligned}
\psi_{\Lambda}=\phi_{0 \Lambda}e^{-\mathcal{S}^{4 \hspace{1 mm} \beta_-=0}_{(+0)}}
\end{aligned}
\end{equation}

\begin{figure}[!ht]
\begin{minipage}[c]{0.4\linewidth}
\includegraphics[scale=.083]{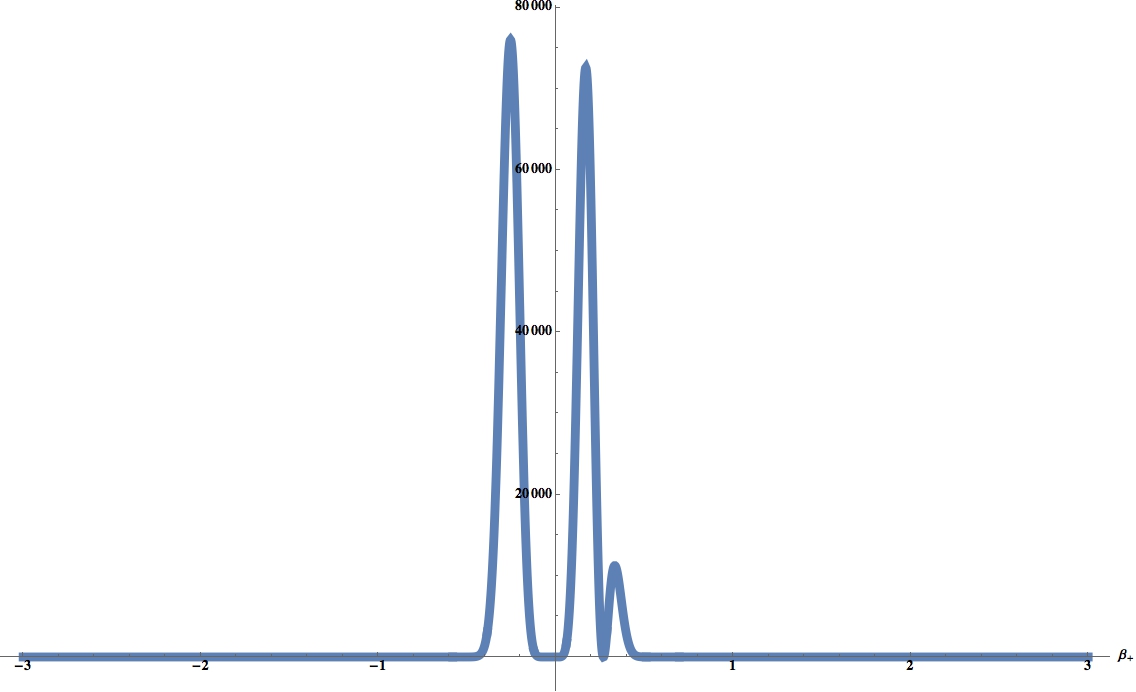}
\caption{ $\alpha=1.3$ \hspace{1 mm} $\Lambda=-1$ \hspace{1 mm} b=0}
\end{minipage}
\hfill
\begin{minipage}[c]{0.4\linewidth}
\includegraphics[scale=.083]{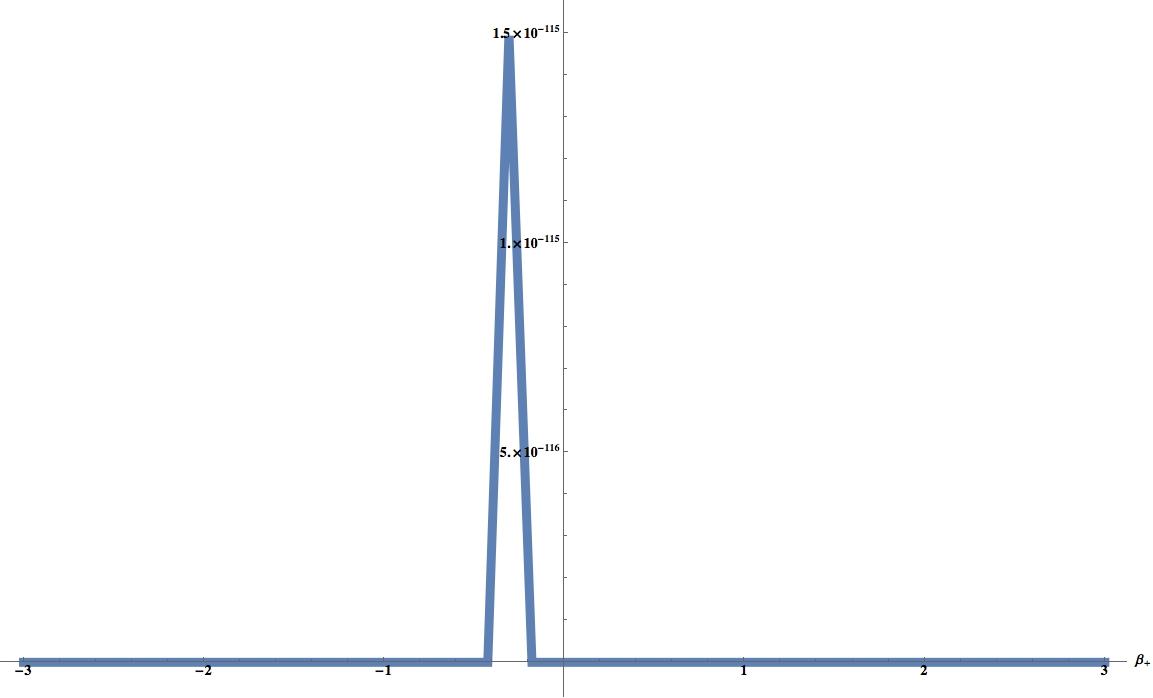}
\caption{ $\alpha=1.3$ \hspace{1 mm} $\Lambda=-1$ \hspace{1 mm} b=2}
\end{minipage}%
\end{figure}

\begin{figure}[!ht]
\centering
\includegraphics[scale=.083]{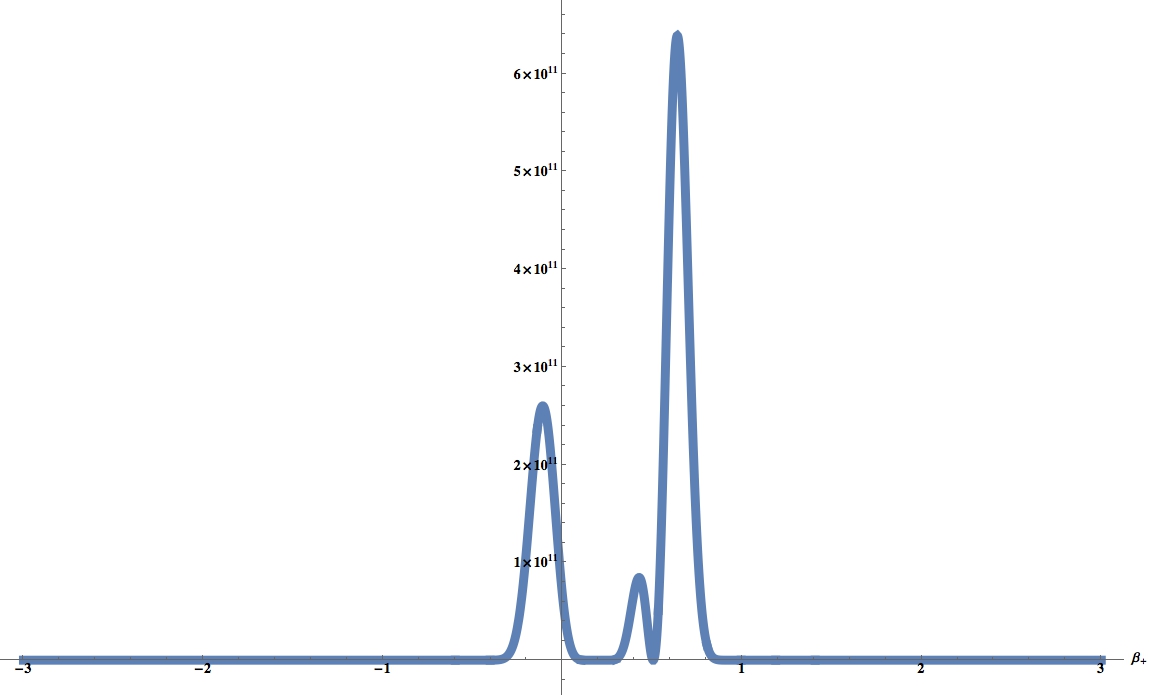}
\caption{ For $\alpha=1.3$ \hspace{1 mm} $\Lambda=0$ \hspace{1 mm} b=2  \hspace{1 mm} }
\label{fig:my_label}
\end{figure}

\section{\label{sec:level1}Discussion}
To begin analyzing our results we first need to adopt an interpretation for the wave functions we computed in this work. Two interpretations of quantum mechanics which in the past have been used to extrapolate physics from Wheeler Dewitt wave functions within the context of quantum cosmology are the consistent histories approach \cite{griffiths1984consistent} and the pilot wave approach \cite{bohm1952suggested}. However, for our purposes we will use the following admittingly naive interpretation which we will briefly outline. Even though we  cannot interpret $|\psi|^{2}$ as a probability density due to the lack of a known dynamical unitary operator, if we fix $\alpha$, and only consider wave functions whose leading order terms don't introduce the pathologies previously mentioned our wave functions are reminiscent of normalizable probability densities as can be seen from our plots. Each point in those plots at a fixed $\alpha$ represents a potential geometric configuration the universe can possess which is specified by the values of the Misner variables $\left(\alpha,\beta_+,\beta_-\right)$. Associated with each of those points in $\beta$ space at a fixed $\alpha$ is a value of $|\psi|^{2}$; it is not unreasonable to conjecture that the greater the value of $|\psi|^{2}$ is, the more likely a Bianchi IX universe will possess the geometry given by the $\beta_+$ and $\beta_-$ Misner variables. For example if $|\psi\left(\alpha,\beta1_+,\beta1_-\right)|^{2} > |\psi\left(\alpha,\beta2_+,\beta2_-\right)|^{2}$ we would interpret this to mean that a Bianchi IX universe described by $\psi$ when it reaches a size dictated by $\alpha$ is more likely to have a spatial geometry which possesses a level of distortion from a perfectly round $S^3$ described by the values of $\left(\beta1_+,\beta1_-\right)$ as opposed to $\left(\beta_2+,\beta_2-\right)$. A problem with our given interpretation though is that it breaks down when the wave function of the universe goes off to infinity when either $\beta_+$ or $\beta_-$ grows without bound, as is the case for Wheeler DeWitt wave functions whose leading order terms are the pathological ones we previously mentioned. Also our interpretation breaks down if the magnitude of the wave function goes off to infinity as $\alpha$ approaches infinity like in the 'no boundary' case. 

Another problem is that it cannot assign numerical values of probability to a micro ensemble of geometric configurations because whatever normalizable constant one uses at a fixed $\alpha$ will not keep the distribution normalized for a different value of $\alpha$. This interpretation is best at analyzing wave functions which don't grow without bound as any of the Misner variables approach positive or negative infinity. Adopting this problematic interpretation is not the author's endorsement of this approach over others for understanding results in quantum cosmology. Rather, we are picking this interpretation because it is intuitive for the solutions we are dealing with and facilitates the elucidation of the points the author wishes to make. In essence the author was inspired to pick this interpretation despite all of its shortcomings because he would like to let the bare solutions speak for themselves. The author strongly encourages future work to be done in extrapolating physics for the results presented in this paper using both the Bohemian approach and consistent histories.

A noteworthy feature of the closed form solutions (43) we found is that as $\alpha$ approaches $-\infty$, which physically corresponds to a universe that is approaching a spatial singularity, our wave functions approach zero. In light of the interpretation we are using this means that the Bianchi IX universes described by our wave functions (43) are forbidden to form singularities in the same way a particle in an infinite potential well is forbidden to tunnel out of it because its wave function vanishes at the boundary. One may be inclined to say that these wave functions represent universes which are singularity free, and perhaps the key to resolving the singularity problem in contemporary cosmology is to introduce anistropic degrees of freedom which played a large role in the early universe but rapidly became less important after inflation. The author would strongly discourage interpreting the vanishing of our closed form solutions this way. The full functional Wheeler DeWitt theory of quantum gravity itself probably isn't an accurate description of nature near the Planck length \cite{calcagni2017classical}. This is due to the fact that it has the spatial metric operator as its configuration variables. We have good reasons to believe that at the Planck scale the description of space-time as a smooth manifold with a Lorentzian signature metric breaks down \cite{bronstein2012republication} \cite{bronstein1936quantentheorie}. Because the Wheeler DeWitt theory comes directly from promoting the spatial metric, which is thought to break down at the Planck scale to an operator, it is unlikely to give an accurate description of nature near the so called singularity. Furthermore, we didn't even solve the functional Wheeler DeWitt equation of quantum gravity, rather by performing symmetry reduction we reduced the degrees of freedom of the classical theory, and as a result obtained a quantum theory of cosmology with only fintely many degrees of freedom. Thus, our results are only merely an approximation of a particular quantum theory of gravity, which itself is not a good description of nature near the singularity. At best our results capture some of the qualitative features that quantum gravity has on the development of a universe which is much larger than the Planck volume, but also small enough so that quantum mechanical effects cannot be ignored \cite{calcagni2017classical}.

Another interesting feature of our wave functions (43) is that they vanish either when $\beta_-$ is zero, or when both $\beta_-$ and $\beta_+$ are zero, thanks to the first order quantum corrections $\mathcal{S}_{(1)}$ they all possess. As $\alpha$ approaches infinity or as the size of the universe grows without bound, these wave functions do not peak at an isotropic(origin of $\beta$ space) geometry like the Bianchi IX solutions for ordering parameters $B= \pm 6 $ do \cite{bae2014quantizing}. Mathematically this can be seen because the Sinh term which is a function of the betas is multiplied to the exponential portion of our wave function, and is independent of $\alpha$. If we interpret these wave functions as 'ground' states with a first order $\mathcal{S}_{(1)}$ quantum correction, we can see how residual artifacts originating from certain quantum effects that were prominent in the early universe in these Bianchi IX universes persist no matter how large they grow. This is a potential example of how features of a large universe, most notably residual anisotropy can have quantum mechanical origins. 

If we look at semi-classical approximate solutions, Figure 3 of our superposition of Joseph Bae's solutions can be interpreted to mean that at $\alpha=0$, this particular universe can tunnel between four different geometries. Three of these geometries are centered away from the origin(isotropy), and are due to the 'excited' states, while the highest peak originates from the 'ground' state. However as $\alpha$ grows further the overall magnitude of the wave function decreases, but the peaks originating from the 'excited' states completely overtake the peak from the 'ground' state as can be seen in figure 4. This would seem to indicate for the 'wormhole' boundary condition, that if a Bianchi IX universe can be described as a superposition of 'ground' and 'excited' states, and if it grew very large it most likely would $\boldsymbol{not}$  be peaked at isotropy.  

Our 'no boundary' and "arm3" wave functions restricted to the $\beta_+$ axis can be interpreted similarly to the superposition of Joseph Bae's 'wormhole' solutions. Our superposition of semi classical states with different boundary conditions give us an interesting picture. For just the "arm" solutions in figure 8 at $\alpha=0$ the wave function isn't strongly peaked at a particular geometry. Its highest peak is centered around isotropy but it also has three ridges of roughly equal magnitude which form a continuum of possibly many anisotropic states the Bianchi IX universe can be in. If we let the plot in figure 9 speak for itself, it appears to be indicating that as $\alpha$ grows the Bianchi IX universe with "arm" boundary conditions settles in an anisotropic state. Unlike our previous wave functions which were peaked at anistropy, these "arm" solution are 'ground' states which were formed solely from the $\mathcal{S}_{(0)}$ term. Because of their tendency as $\alpha$ grows to not peak at isotropy, another proper name of these "arm" solutions would be the "non-isotropic" or "anistropic" solutions. Figure 10 shows that depending on how the weights between the 'excited' 'wormhole' states and "arm" states are assigned, that tunneling between sharply peaked geometric states is still possible, but there will be a lot more noise in terms of the possible configurations the universe can have due to the continuous ridges of the "arm" solutions. In figure 11 we see that the ridges extend further away from the origin than the peaks belonging to the 'excited' states with 'wormhole' boundary conditions. As a result for large $\alpha$ the "arm" solutions dominate the behavior of the wave function, and as previously mentioned have a great affinity for driving the Bianchi IX universe to a highly anistropic state. If we include the 'no boundary' solutions we see that they dominate the wave function for large $\alpha$. However, because the magnitude of $\psi$ grows without bound as $\alpha$ approaches infinity it is unclear how to form a cogent interpretation of what is occurring physically in a universe represented by that wave function (figure 12 and 13).

By examining our leading order solutions with $\Lambda$ it can be seen that a cosmological constant is a driver for anistropy for the Bianchi IX models. This is reflected in the fact that as $\alpha$ approaches infinity, the leading order solutions do not peak at isotropy as can be seen in figure 15, a residual level of anistropy remains for anti de Sittter $(\Lambda < 0)$ universes.  Our leading order $\mathcal{S}^i_{\pm 0)}$ terms indicate that the effects of the cosmological constant become most prominent in quantum cosmology for large $\alpha$. This is not surprising considering the cosmological constant is thought to be related to what we call dark energy \cite{peebles2003cosmological}; and as the size of the universe increases so does the amount of dark energy in it. Unsurprisingly for $\alpha$ $\ll$ 0 the effects of the cosmological constant are practically negligible because the term proportional to $\Lambda$ in $\mathcal{S}^i_{\pm 0)}$ decays as $e^{4\alpha}$. Even though what we may have said can appear obvious to the reader, it is important to check that features of the classical system manifest themselves in its quantum analog and in the case of the addition of the cosmological constant to the vacuum diagonalized Bianchi IX models they do. If we include a primordial magnetic field we obtain what we would expect classically. Notably that the primordial magnetic field has a strong effect when the universe was small but becomes almost negligible when it is large due to the primordial nature of the field. As can be seen in figure 16 a very strong primordial field can greatly increase early universe anistropy, but unlike the cosmological constant for large $\alpha$ its effects rapidly diminish. For the solutions restricted to the $\beta_+$ axis when both a magnetic field, and a cosmological constant are present, the addition of an magnetic field can decrease the amount of potential geometric configurations a Bianchi IX universe can take on as can be seen by comparing figures 18 and 19. 

The above discussion isn't meant to be a rigorous dissection of what is truly going on. But rather a very simple surface analysis of the results presented in this paper. 

\section{\label{sec:level1}Conclusion}
The diagonalized Bianchi IX symmetry reduced Wheeler DeWitt equation has been studied for over 50 years since Misner \cite{misner1969quantum} first penned it down. Closed form solutions of any kind have been very hard to come by. In this work we significantly expand the number of known closed form solutions to the Bianchi IX symmetry reduced Wheeler DeWitt equation for two particular Hartle-Hawking ordering parameters by applying a Euclidean-signature semi classical method, and constructing our $\mathcal{S}_{(1)}$ term so that they have as many non trivial free parameters as possible; in the hopes that those parameters can be tuned to truncate the infinite sequence of equations this semi-classical method generates. Furthermore we analyzed Bianchi IX wave functions restricted to the $\beta_+$ axis corresponding to the 'no boundary' boundary condition and for an "arm" boundary condition. We were able to construct some interesting semi-classical 'excited' states from these restricted wave functions. Finally we were able to obtain new solutions for the Eucldiean-signature Hamilton Jacobi equation for the Bianchi IX model with a cosmological constant, and a primordial magnetic field present. These solutions were used to construct semi-classical leading order solutions to the Lorentizan signature symmetry reduced Wheeler DeWitt equation. These results showcase how powerful Euclidean-signature semi classical methods are for solving problems in quantum cosmology.

The work presented in this paper can be extended to non communicative \cite{maceda2008homogeneous} \cite{garcia2002noncommutative} Bianchi IX quantum cosmology. A through analysis of these new solutions using the consistent histories \cite{griffiths1984consistent} \cite{craig2011consistent} or Bohmnian approach \cite{bohm1952suggested} \cite{pinto2013quantum} can shed much light on the salient features of Bianchi IX quantum cosmology. Similarly the work in this paper can further explored in a supersymmetric setting. Further work can be carried out to derive a globally defined  ${\mathcal{S}_{(1)}}$ term for the case when a cosmological constant is present which can potentially open the doors to penning a closed form solution to the actual Bianchi IX Wheeler Dewitt equation, or at least proving that one exists.

Prior to writing this paper the author applied this semi
classical method, and other related methods to obtains
solutions to the Lorentzian signature symmetry reduced
Wheeler DeWitt equation for other Bianchi A models,
and for the LRS Bianchi IX and VIII models with and
without cosmological constant. In addition the author
applied this semi classical method to the case of a rotating Bianchi IX universe with matter and extended the results of [65] to Bianchi VIII and VII universes with a cosmological constant, and went beyond leading order. The author plans to publish these findings in due time. The potential applications of the Euclidean-signature semi-classical method to problems in quantum cosmology and beyond are quite vast. As a result the author very much looks forward to seeing what future applications of this
method will produce.

\section{\label{sec:level1}ACKNOWLEDGMENTS}
 
I am grateful to Professor Vincent Moncrief for valuable discussions at every stage of this work. I would also like to thank George Fleming for facilitating my ongoing research in quantum cosmology. Daniel Berkowitz acknowledges support from the United States Department of Energy through grant number DE-SC0019061. I also must thank my aforementioned parents.

\bibliography{Bianchi_2}

\end{document}